\documentclass[a4paper,fleqn]{cas-dc}

\usepackage[authoryear]{natbib}
\def\tsc#1{\csdef{#1}{\textsc{\lowercase{#1}}\xspace}}
\tsc{WGM}
\tsc{QE}
\tsc{EP}
\tsc{PMS}
\tsc{BEC}
\tsc{DE}
\begin{document}
\let\WriteBookmarks\relax
\def\floatpagepagefraction{1}
\def\textpagefraction{.001}

\shorttitle{HE: An Analysis of its Applications in SE}

\shortauthors{Amorim et~al.}

\title [mode = title]{Homomorphic Encryption: An Analysis of its Applications in Searchable Encryption}                      
%\tnotemark[1,2]

% \tnotetext[1]{This work was partially supported by the Norte Portugal Regional Operational Programme (NORTE 2020), under the PORTUGAL 2020 Partnership Agreement, through the European Regional Development Fund (ERDF), within project ”Cybers SeC IP” (NORTE-01-0145-FEDER-000044).}

%  \tnotetext[2]{This work was partially supported by the Norte Portugal Regional Operational Programme (NORTE 2020), under the PORTUGAL 2020 Partnership Agreement, through the European Regional Development Fund (ERDF), within project ”Cybers SeC IP” (NORTE-01-0145-FEDER-000044).}

\author[1]{Ivone Amorim}[type=editor,
                        auid=000,bioid=1,
                        %prefix=Ms,
                        %role=Researcher,
                        orcid=0000-0001-6102-6165]

\cormark[1]


\ead{ivone.amorim@portic.ipp.pt}

%\ead[url]{www.cvr.cc, cvr@sayahna.org}

%\credit{Conceptualization of this study, Methodology, Software}

\affiliation[1]{organization={PORTIC - Porto Research, Technology and Innovation Center, Polytechnic Institute of Porto (IPP)},
    %addressline={}, 
    %city={Porto},
    % citysep={}, % Uncomment if no comma needed between city and postcode
    postcode={4200-374, Porto}, 
    country={Portugal}}

%Polytechnic Institute of Porto (IPP)

\author[1]{Ivan Costa}[style=chinese]

% \author[2,3]{CV Rajagopal}[%
%    role=Co-ordinator,
%    suffix=Jr,
% \ead{cvr3@sayahna.org}
% \ead[URL]{www.sayahna.org}

%\credit{Data curation, Writing - Original draft preparation}

% \affiliation[2]{organization={Sayahna Foundation},
%     % addressline={}, 
%     city={Jagathy},
%     % citysep={}, % Uncomment if no comma needed between city and postcode
%     postcode={695014}, 
%     state={Trivandrum},
%     country={India}}

% \author%
% [1,3]
% {Rishi T.}
% \cormark[2]
% \fnmark[1,3]
% \ead{rishi@stmdocs.in}
% \ead[URL]{www.stmdocs.in}

% \affiliation[3]{organization={STM Document Engineering Pvt Ltd.},
%     addressline={Mepukada}, 
%     city={Malayinkil},
%     % citysep={}, % Uncomment if no comma needed between city and postcode
%     postcode={695571}, 
%     state={Trivandrum},
%     country={India}}

\cortext[cor1]{Corresponding author}
%\cortext[cor2]{Principal corresponding author}

% \fntext[fn1]{This is the first author footnote. but is common to third
% \fntext[fn2]{Another author footnote, this is a very long footnote and
%   it should be a really long footnote. But this footnote is not yet
%   sufficiently long enough to make two lines of footnote text.}

%\nonumnote{This work was partially supported by the Norte Portugal Regional Operational Programme (NORTE 2020), under the PORTUGAL 2020 Partnership Agreement, through the European Regional Development Fund (ERDF), within project ”Cybers SeC IP” (NORTE-01-0145-FEDER-000044).}

\begin{abstract}
The widespread adoption of cloud infrastructures has revolutionized data storage and access. However, it has also raised concerns regarding the privacy of sensitive data stored in the cloud. To address these concerns, encryption techniques have been widely used. However, traditional encryption schemes limit the efficient search and retrieval of encrypted data. 
To tackle this challenge, innovative approaches have emerged, such as the utilization of Homomorphic Encryption (HE) in Searchable Encryption (SE) schemes. This paper provides a comprehensive analysis of the advancements in HE-based privacy-preserving techniques, focusing on their application in SE. The main contributions of this work include the identification and classification of existing SE schemes that utilize HE, a comprehensive analysis of the types of HE used in SE, an examination of how HE shapes the search process structure and enables additional functionalities, and the identification of promising directions for future research in HE-based SE. The findings reveal the increasing usage of HE in SE schemes, particularly Partially Homomorphic Encryption. The analysis also highlights the prevalence of index-based SE schemes using HE, the support for ranked search and multi-keyword queries, and the need for further exploration in functionalities such as verifiability and the ability to authorize and revoke users. Future research directions include exploring the usage of other encryption schemes alongside HE, addressing omissions in functionalities like fuzzy keyword search, and leveraging recent advancements in Fully Homomorphic Encryption schemes.
\end{abstract}

% \includegraphics{figs/grabs.pdf}

% Keywords

\begin{keywords}
Searchable Encryption \sep Homomorphic Encryption \sep Secure Search \sep Data Privacy \sep Keyword Search
\end{keywords}

\maketitle

\section{Introduction}\label{introduction}
The rapid growth and widespread adoption of cloud infrastructures have revolutionized the way we store and access information. From personal data backup and file-sharing services like Dropbox and Google Drive to enterprise-level data management and scalable infrastructure solutions, cloud storage has had a profound impact on our day-to-day lives, particularly in sectors such as Healthcare (\citealt{8777606}; \citealt{Ghosh20141}) and Education (\citealt{Agrawal2021}; \citealt{Gonzalez-Martinez2015132}). According to a report published by \citet{netwrix_cloud_2022}, 80\% of organizations store sensitive data in the cloud. The advantages it offers include high data availability, convenient access from anywhere, reduced infrastructure costs, unlimited storage space, and cost-effectiveness (\citealt{Yang2020DataSA}).
However, with cloud adoption come heightened security concerns as sensitive information entrusted to cloud servers faces potential vulnerabilities, such as unauthorized access, data breaches, and insider threats. Netwrix's report~\citeyearpar{netwrix_cloud_2022} also concluded that 55\% of healthcare organizations had experienced a third-party data breach in the last 12 months, which was the second highest percentage of all industry sectors, beaten only by the financial sector where 58\% of companies had experienced a third-party data breach. Both of these industry sectors rely heavily on third parties, and those third parties have access to sensitive data, which is of high value to cybercriminals. Moreover, their results show that attacks have become more sophisticated and harder to spot. 
To address these issues, ensuring the privacy of users' sensitive data becomes crucial. The most common way to achieve privacy is through encryption, where data is encrypted before being stored in the cloud, providing end-to-end data privacy. However, traditional encryption schemes pose challenges when it comes to efficiently searching and retrieving data stored in encrypted form, limiting the usability of encrypted storage solutions. Naive approaches, such as downloading all ciphertexts, decrypting them, and searching on plaintexts, are impractical. Therefore, to tackle these challenges, innovative concepts such as Secure Search, Private Information Retrieval~(PIR), and Searchable Encryption~(SE) have emerged. Secure search addresses the need to perform search operations while maintaining the privacy and confidentiality of the data (\citealt{akavia_secure_2018}). PIR, closely related to secure search, enables users to retrieve specific data from a database without revealing which items they are accessing (\citealt{10.1155/2021/5553256}). SE, on the other hand, is a cryptographic technique that allows for secure and efficient search operations over encrypted data (\citealt{sharma_searchable_2023}). It is important to note that the terms ``Secure Search'' and ``Searchable Encryption'' are sometimes used interchangeably, blurring the distinction between the two concepts. In practice, the boundaries between these terms can be fluid, with different researchers and practitioners using them to refer to related techniques and mechanisms.
Additionally, PIR is often considered a component within secure search or SE approaches, as it enables private data retrieval from a database. In this work, we mainly use the term ``Searchable Encryption'' while ``Secure Search'' is used whenever we feel a distinction is needed. 

Within the realm of SE, one notable approach is the utilization of Homomorphic Encryption~(HE). HE is a special type of encryption that allows computation on encrypted data without decryption (\citealt{acar_survey_2018}). This property makes it well-suited for privacy-preserving techniques, enabling search operations directly on encrypted data while maintaining confidentiality. Researchers have increasingly focused on this approach, as it offers an attractive framework for secure search without requiring costly setup procedures~(\citealt{choi_compressed_2021}).
Numerous approaches have been proposed that leverage the homomorphic properties to provide privacy-preserving techniques, particularly in the context of Searchable Encryption. However, there is a lack of dedicated studies analysing the advancements in this area. On the other hand, it is crucial for the scientific community to gain a comprehensive understanding of the current state-of-the-art and identify promising directions for further exploration in this subject.

In this work, we aim to provide a comprehensive analysis of the advancements in HE-based privacy-preserving techniques, specifically focusing on their application in SE. 

\subsection{Related Work}
Since the publication of the first work on SE, due to \citet{song_practical_2000}, several surveys and review papers have been proposed on this topic, which highlights the growing importance of secure data management in a world where the majority of data is stored in third-party cloud services.

\citet{bosch_survey_2014} presented the first survey on SE. In this work, a comprehensive discussion is given with respect to the number of writers and readers supported by each scheme. The authors consider four different cases are considered: single-writer/single-reader, multi-writer/single-reader, single-writer/multi-reader, and multi-writer/multi-reader. The schemes are organized based on their query expressiveness, and a comparative analysis is made in terms of efficiency and security. \citet{wang_secure_2016} and \citet{han_secure_2016} published two new systematic surveys on SE. The former gives a broader view of the existing SE schemes based on their usage in symmetric or asymmetric key settings, while the latter categorizes the systems according to three aspects: security requirements, search functionalities, and deployment model. Moreover, as stated by the authors, the latter also introduces a new deployment model which was not covered in the work of \citet{bosch_survey_2014}, called the Server-User model, in which the cloud server is the owner of the data and acts as both the data owner and storage server.

In later years, other works have been published that provided reviews on existing and new SE methods~(\citealt{dowsley_survey_2017}, \citealt{poh_searchable_2017}, \citealt{pham_survey_2019}, \citealt{handa_searchable_2019}).  

More recently, \citet{andola_searchable_2022} published a comprehensive review paper that focuses on analysing the features and limitations of these techniques based on their performance and robustness against various types of attacks. One of the key aspects of this work is the in-depth analysis of each technique and the cryptographic basis that determines their efficiency. Also in the same year, \citet{noorallahzade_survey_2022} published a complete classification of SE schemes based on a comprehensive set of metrics, including search type, index type, results type, security models, type of implementation, the multiplicity of users, cryptographic primitives and technique used. For each category, the available schemes were compared and evaluated. \citet{sharma_searchable_2023} provided a comprehensive guide on SE for non-security experts. The main goal of this work was to help general practitioners select the most suitable SE scheme for their specific needs by presenting a survey that details the existing schemes based on five key characteristics: key structure, search structure, search functionality, support for readers/writers and reader capability.  Both Symmetric and Asymmetric SE schemes are discussed. It also presents a comparative analysis, which may assist non-security experts in making informed decisions about their encryption needs. 

Over the years, other works have been published that survey the application of SE in different contexts. For example, \citet{zhang_searchable_2018} discussed the use of SE in healthcare systems, while \citet{bader_searchable_2021} focused on its application in the industrial Internet of Things (IoT).  Other works have also analysed how new technologies such as Blockchain are being used to enhance the potential of SE mechanisms~(\citealt{how_blockchain-enabled_2022}, \citealt{pillai_blockchain-based_2022}). However, no study is devoted to the use of HE in secure search mechanisms. In this context,  our main objective is to survey SE techniques that incorporate HE into their design, while also providing a comprehensive analysis of how HE is used and what benefits it brings.

\subsection{Main Contributions}
 The main contributions of our work are:
\begin{enumerate}
    \item To identify and classify existing SE utilizing HE. Our analysis covers several aspects of these systems, including the encryption techniques used, the structure of the search process (sequential scan or index-based), and the search capabilities offered (such as the ability to handle multiple keywords, regular expressions, wildcards, phrases, ranges, occurrences, and fuzzy keywords). Additionally, we study other functionalities like authorization and access revocation, static and dynamic approaches to SE, and correctness verification.
    \item To conduct an extensive analysis of the most common types of HE used in SE schemes, focusing our analysis on the identification of whether the schemes employed fall under the categories of Partially Homomorphic Encryption, Somewhat Homomorphic Encryption, or Fully Homomorphic Encryption. Furthermore, we explicitly identify the HE schemes whenever possible.
    \item To examine how HE is used to achieve the different properties and characteristics within SE, building upon the categorization mentioned earlier. Specifically, we investigate how HE shapes the search process structure, enhances search capabilities, and enables additional functionalities in SE.
    \item To identify promising directions for future research and development in HE-based SE schemes, aiming to deliver more flexible and advanced solutions for SE.
 
\end{enumerate}

\subsection{Research Methodology}\label{ResearchMethodology}
To perform a comprehensive analysis of the application of HE in SE, a systematic approach was adopted. This approach involved searching several academic databases, including Elsevier ScienceDirect, Scopus, ACM Digital Library, IEEE Xplore, and Web of Science.

To conduct the search in the previously mentioned databases, a list of keywords was identified after thoroughly reviewing the main literature on this area. The listed databases were searched for works that included at least one keyword related to SE and one related to HE in their title, abstract, or set of keywords. Keywords related to SE included ``searchable'', ``secure search'' and ``keyword search''. Keywords related to HE included ``homomorphic'' and ``encrypted''. 

%As such, the main research query used was

%\begin{center}\emph{
%    (searchable OR ``secure search'' OR 
%``keyword search'') AND 
%(homomorphic OR encrypted)}
%\end{center}

After conducting an initial search, a screening process was performed to select only publications that met the following criteria: they were written in English, published in peer-reviewed journals or conferences after 2016, and focused on secure search methods or SE schemes that leverage the properties of HE in the search process.

The search was conducted in February 2023 and {\bf 290} distinct papers were identified. Out of these, {\bf 23} were determined to meet the established inclusion criteria.

\subsection{Organization}
This paper is organized as follows:
 Section~\ref{SearchableEncryption} introduces the main actors and processes involved in SE schemes, as well as how they can be characterized in terms of Search Structure, Search Functionalities, Multiplicity of users, and Other functionalities. The definition of HE schemes, the main existing types, and approaches are presented in Section~\ref{HomomorphicEncryption}. The comprehensive analysis of the selected works is presented in Section~\ref{sec:analysis}, while the discussion of the research trends is covered in Section~\ref{sec:trends}. Finally, in Section~\ref{sec:conclusion} section, we provide our conclusions and suggest research directions.

\section{Searchable Encryption}\label{SearchableEncryption}
SE refers to a cryptographic technique that enables searching over encrypted data without the need to decrypt it. There are two main categories of SE techniques based on the underlying encryption process: symmetric and asymmetric. Similarly to conventional symmetric encryption schemes, symmetric SE techniques typically use the same key for encryption and decryption, while in asymmetric SE schemes, a pair of keys is used - one for encryption and one for decryption. 

\begin{figure}[t]
    \centering
     \includegraphics[width=0.95\columnwidth]{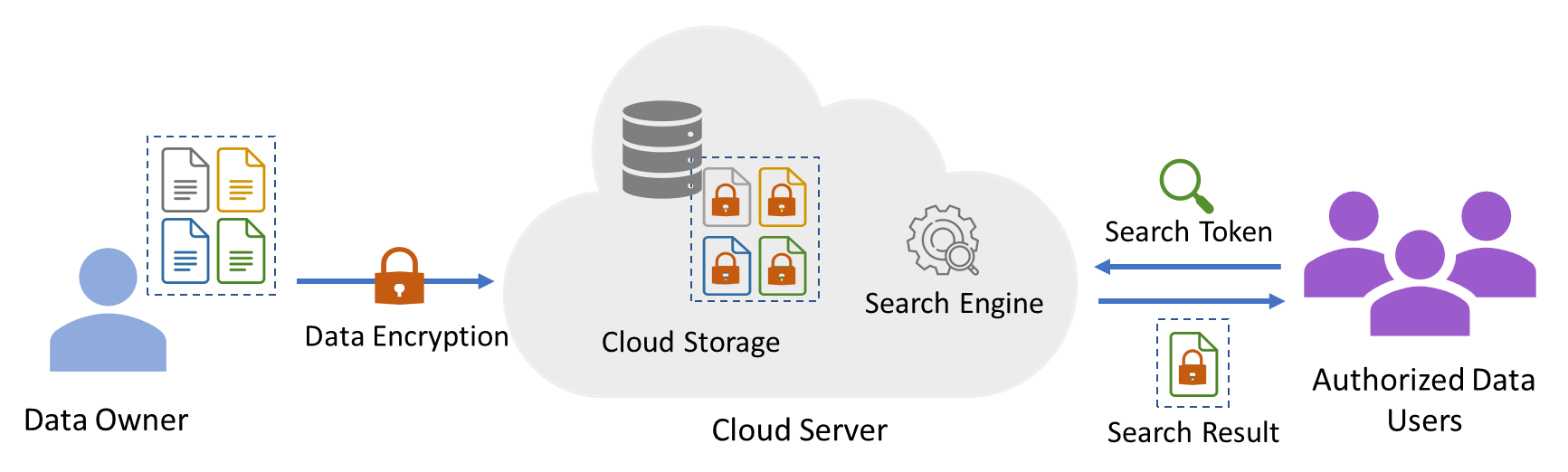}
    \caption{Conceptual overview of a cloud-based SE system.}
    \label{Figure1}
\end{figure}

Figure~\ref{Figure1} depicts a high-level architecture of a generic cloud-based SE system which is usually composed of three main entities, as follows:

\begin{itemize}
    \item {\bf Data Owner (DO)} The data owner (sometimes referred to as ``client'' in relevant literature, \citealt{sharma_searchable_2023}) is the entity responsible for encrypting the data and outsourcing it to the cloud server. Generally, the DO is the producer of the data and has legitimate control and ownership over it. However, in some scenarios, another entity, known as the {\bf data provider}, may be responsible for generating the data.
    The SE methods that use indices to aid in the search process, also known as index-based SE schemes, typically require the data provider (whether it is the data owner or not) to encrypt the index and share it with the cloud.
    
    \item {\bf Data User (DU)} A data user represents an entity that wishes to search over the DO's encrypted data. Ideally, it is able to perform the search only if previously authorized by the DO. The DU is, therefore, responsible for sending a search request to the cloud server which, after processing the request, returns the results. It is worth noting that, in some systems, the DO can also act as a DU.
   
    \item {\bf Cloud Server (CS)} The cloud server is the entity responsible for securely storing the encrypted data and providing a SE service to authorized DUs. It receives the encrypted documents uploaded by the DO and carries out three main tasks: storing the data, searching the data, and maintaining the search data structures updated. The way the CS performs the search depends on the SE scheme being used. In the case of an index-based SE, the cloud server performs the search by comparing the search request with the secure index, and it then sends the results to the authorized DU. However, for SE schemes that are not index-based, the process of searching the encrypted data usually requires scanning the whole document, as will be discussed later in this chapter.
\end{itemize}

The architecture of a SE system usually includes four processes involving the previously described entities. The specific algorithms used in each step may differ based on the specific design and requirements of the SE scheme. 

The following description of each process has been made as broad as possible, taking into account various factors such as whether an index-based search mechanism is present or not, and whether it is a symmetric or asymmetric SE scheme.

\begin{enumerate}
    \item \textbf{Setup()} Based on an input security parameter $\lambda$, the algorithm generates various system parameters $P$ as well as the necessary keys $K$. In asymmetric SE, a public and a private key are generated, while in symmetric SE, only a single key is needed. This algorithm is usually executed by the system owner.

    \item \textbf{Encryption()} An encryption algorithm encrypts the data using the key(s) generated in the previous process K, and it outputs a message $M'$ obtained by encrypting the original data $M$ using an encryption algorithm $E$.
    
    If the SE is index-based, this algorithm will also use the input key(s) $K$ to encrypt an input set of keywords $W$, which may be used to generate an Index $I$ of encrypted keywords. Then, the DO uses this algorithm to construct a Searchable Ciphertext ($SC$) by associating the encrypted Index $I$ with the encrypted message $M'$. The SC can then be uploaded to the cloud server. 
    
    If the SE method is scan-based (meaning that the server will scan the encrypted data directly), the previous step can be skipped, and the message $M'$ can be stored directly on the cloud server without generating an SC.

    \item \textbf{TokenGen()} This algorithm, sometimes also called Trapdoor, is used by authorized users to generate search queries. 
    It takes an input encryption key $K$ and an input query $Q={k_1, k_2, ..., k_m}$ and generates a search token $T_Q$. The specific implementation of the algorithm depends on whether the SE scheme is index-based or scan-based.
    For an index-based scheme, the algorithm encrypts each keyword in the query $Q$ using the encryption key(s) $K$ and generates an index $I={w_1', w_2', ..., w_m'}$ of encrypted keywords. The search token $T_Q$ is then constructed from $I$.
    For a scan-based scheme, the algorithm  generates a scan token $T_S$. The search token $T_Q$ is then constructed from the chosen search query $Q$ and the scan token $T_S$. 
    Once constructed, the search token $T_Q$ is sent to the server, which uses it to search for relevant data in the encrypted database.

    Depending on the scenario, the query may only be performed by the DU. 

    \item \textbf{Search()} The search algorithm is used by the CS to search the encrypted data for matches to a search query. 
    In an index-based scheme, the search algorithm applies the search token $T_Q$ onto the searchable ciphertext $SC$ to identify the set of encrypted keywords, and corresponding indices, that match the query. Then, it retrieves the corresponding encrypted data to the DU.
    In a scan-based scheme, the search algorithm applies the search token $T_Q$ and the scan token $T_s$ onto the encrypted data to identify the set of searchable data segments having keywords matched with the keywords of the search query. Once the matches are found, the server sends the search result to the Data User.  
\end{enumerate}

\subsection{Characterization of a SE scheme}\label{sec:charact}

 SE schemes can be categorized in various ways, and there are several categorizations of SE schemes in the literature. For example, \citet{han_secure_2016} categorized SE systems based on three aspects: security requirements, search functionalities, and deployment model. \citet{noorallahzade_survey_2022} presented a more comprehensive categorization by considering additional aspects such as search type, index type, results type, security models, type of implementation, multiplicity of users, cryptographic primitives and technique used.  \citet{sharma_searchable_2023} detailed existing SE schemes based on five key characteristics: key structure, search structure, search functionality, support for readers/writers, and reader capability.

In our work, we will consider a categorization of SE schemes that includes four categories: Search Structure, Multiplicity of Users, Search Functionalities, and Other Functionalities. By using this categorization, we then analyse each scheme and identify which category it belongs to, allowing us to make a proper analysis of the features that are most common in SE schemes that use HE.

\begin{figure}[t]
    \centering
    \includegraphics[width=0.95\columnwidth]{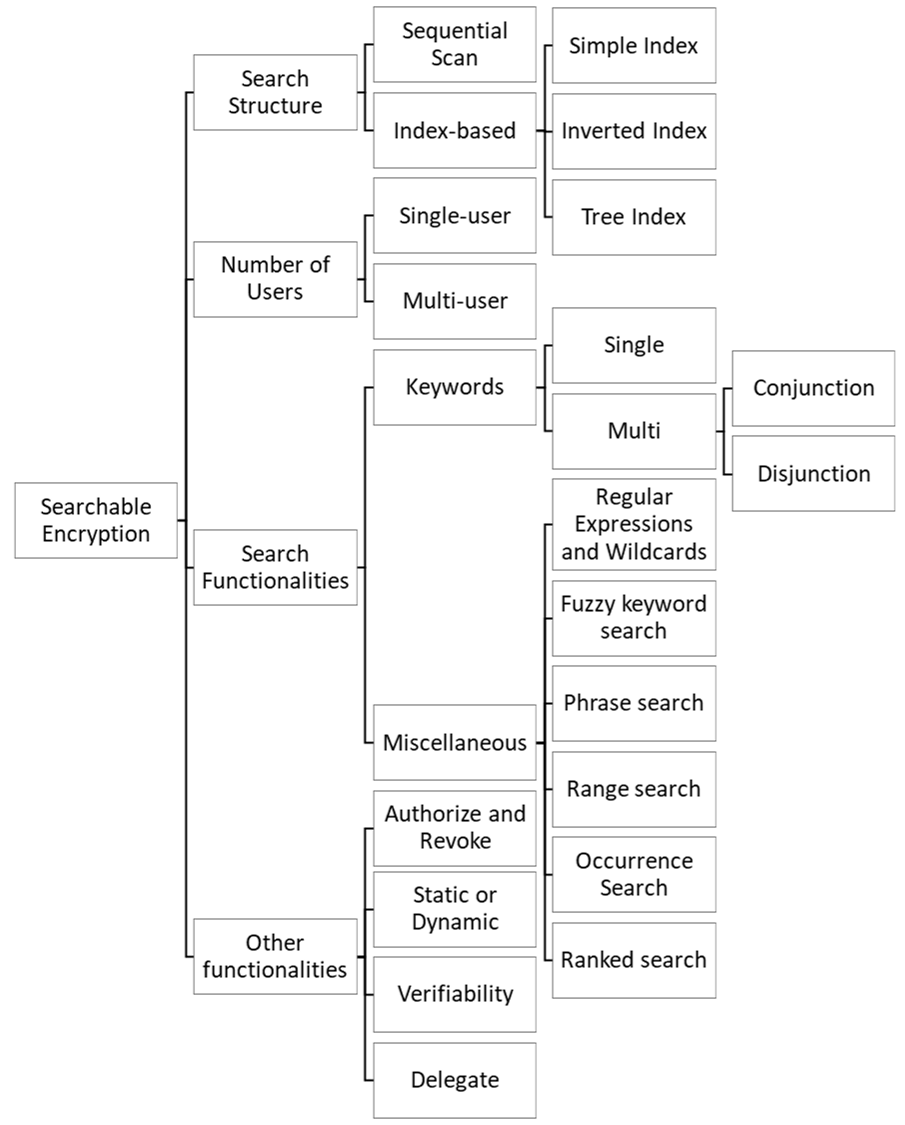}
    \caption{Key characteristics of a cloud-based SE scheme.}
    \label{Figure2}
\end{figure}

The categorization we will use is represented in Figure \ref{Figure2}, and each category is described in detail in the following sections for clarification.

\subsubsection{Search Structure}\label{sec:searchstructure}
 An important aspect of a SE scheme is the search structure used to perform searches over the encrypted data, since it may significantly impact the scheme's efficiency. Historically, the first scheme, due to \citet{song_practical_2000}, used what is called a {\bf sequential scan} search structure. This search operation involves sequentially scanning through all the encrypted documents in the database to find those that match the search query. Although other works in the sequential scan category have been published, such as the one by \citet{boneh_public_2007} on public encryption with keyword search, there are not many prominent research works that follow this approach (\citealt{pham_survey_2019}). This is due to the fact that these techniques are very inefficient and are not well-suited for large databases or frequent search queries, as the search operation can become very time-consuming and computationally expensive. Moreover, this search method is prone to exposing sensitive information to the server, which compromises the privacy and security of the data (\citealt{andola_searchable_2022}). A potential advantage of sequential scan-based SE schemes is that they can provide a simple and straightforward way to search over encrypted data without the need for complex indexing or search structures.

 Another type of search structure commonly used in SE schemes to improve search performance is the {\bf index-based} search structure. In this approach, a special data structure, called Index, is associated with the encrypted documents. This new data structure makes possible to compare the search query with the entries in the index, instead of searching the contents of all the documents in the database. Also, using indexes provides support for working with files of different formats, including compressed and encrypted files, as well as multimedia files (\citealt{handa_searchable_2019}). Nevertheless, with this method, the DO needs to send to the cloud server the encrypted data as well as the encrypted index. These two sets of information can be encrypted with the same encryption scheme, or we can use a SE scheme to encrypt the index and another reliable encryption scheme to encrypt the sensitive data. There are mainly three types of index of keywords: simple index, inverted index, and tree index.
\begin{itemize}
    \item {\bf Simple index} 
    In systems where this approach is used, it is created an index for each document before encrypting and uploading it to the cloud server. The index consists of keywords that are considered relevant to that document. This kind of index is suitable for applications where a small number of documents with a specific value for each keyword field are required to be uploaded to the CS (\citealt{sharma_searchable_2023}).

    \item{\bf Inverted index}
    The term inverted-index comes from the process of building the index backwards.
    This is because, instead of associating each document with a set of keywords, the index is created by coupling each keyword with the set of documents where it appears.  This approach significantly reduces the time required for searching, making it the most suitable search structure for applications that involve uploading a large number of documents to the cloud server (\citealt{sharma_searchable_2023}).
    
    \item \textbf{Tree index}  A tree index is also a very efficient method to optimize the search process. While there are various approaches to building a tree index, the basic idea is to create a tree-like structure containing the searchable keywords, by dividing the set of keywords into smaller sets. When a DU searches for a specific keyword, the CS will search the index-tree, starting at the root and traversing every relevant node until a match is found.
\end{itemize}
  
 Compared to sequential scan approaches, index-based schemes are highly efficient, reducing the comparisons per file from $O(n)$ to $O(1)$.  However, a drawback is the fact that the keywords that can be used in the query are limited to those that were extracted during index generation~(\citealt{handa_searchable_2019}).
%%%%%%%%%%%%%%%%%%%%%%%%%%%%%%%%%%%%%%%%%%%%%%%%%%%%%%%%%
\subsubsection{Multiplicity of Users}
SE schemes allow DOs to securely outsource data storage and DUs to retrieve data based on specific search criteria. Based on the multiplicity of users involved, we further classify these schemes as {\bf single-user}, if the DO assumes the role of DU, and {\bf multi-user} if the DU is different from the DO.
%%%%%%%%%%%%%%%%%%%%%%%%%%%%%%%%%%%%%%%%%%%%%%%%%%%
\subsubsection{Search Functionalities}\label{sec:searchfunc}
The search functionalities of a SE scheme, as the name suggests, determine the kind of queries which can be performed, the filtering and sorting options and other tools that help data users to refine their search. In this work, and following a categorization similar to \citet{sharma_searchable_2023}, we distinguish between search functionalities directly associated with the number of keywords and others. The set of functionalities not directly linked with the number of keywords is referred to as ``Miscellaneous'' (Figure~\ref{Figure2}).

Regarding the number of keywords, a SE scheme can be categorized either as Single or Multi-Keyword. In a {\bf Single-Keyword} SE scheme, the DU is allowed to include just one search term in each query. Consequently, if he/she wants to search for multiple terms using this type of scheme, then he/she needs to perform multiple queries, one for each search term (\citealt{handa_searchable_2019}).
In a {\bf Multi-Keyword} SE scheme, the user is allowed to perform a search with more than one search term in each query. However, it is important to mention that just because multiple keywords are supported, it does not necessarily imply that users can perform searches with an unlimited number of keywords. Some schemes may be restricted to a specific number of keywords.
Furthermore, in a multi-keyword SE scheme, a query can be represented as a simple list of search terms, as a conjunctive query (using the AND relation), or as a disjunctive query (using the OR relation). Depending on the representation of the query, the multi-keyword SE can be categorized as a {\bf Conjunctive Keyword} SE or {\bf Disjunctive Keyword} SE.

Besides the characterization regarding the number of keywords, the existing SE schemes can also be categorized based on miscellaneous functionalities. In this work, we consider the following: Regular Expressions and Wildcards,  Fuzzy keyword search, Phrase search, Range search, Occurrence search, and Ranked search.\\\\

\noindent \textbf{Regular expressions and wildcards}

\noindent SE schemes that allow regular expression or wildcard based search queries are very useful when the DUs know the specific patterns of the keywords that they want to search for in the documents. In both cases, the query can be constructed using special characters to describe the keyword(s) pattern.

If a SE scheme allows wildcard queries, then the DUs can replace a single character in the search query with a symbol known as a wildcard. Generally, two types of wildcards are used: ``?'' and ``*'', the former representing a single character and being called a single-character wildcard, while the latter can represent any number of characters and is called a multi-character wildcard~(\citealt{liu_multi-keyword_2023}). In case a regular expression-based SE scheme is used, a combination of special characters and operators can be employed to describe more complex patterns in search queries than those allowed by wildcards. 
%For instance, if we want to search in a database for all documents containing Portuguese phone numbers which start by $+3519$, we can use a regular expression pattern to describe the structure of those phone numbers. For example, we could query something like $+3519[0-9]\{8\}$ and the cloud would search for every phone number starting by $+3519$.
 While regular expressions provide greater flexibility, wildcard-based SE schemes are a simpler alternative when searching for keywords with minor variations.\\

\noindent\textbf{Fuzzy keyword search}
 
\noindent A typical SE scheme, even when it allows wildcard or regular expression-based search queries, can only search for exact matches of keywords in ciphertexts. It does not allow any typos or inconsistencies in the format of the searched term(s). However, this limitation can be addressed by incorporating the Fuzzy Keyword Search functionality into the SE scheme. With this feature, the scheme is able to search for results related to the correct spell word even if the data users have misspelled it during their search~(\citealt{handa_searchable_2019}).\\

\noindent \textbf{Phrase search}

 \noindent Often, DUs are interested to find specific phrases instead of just individual keywords. Even though this can be achieved using SE schemes that allow for conjunctive search queries, this requires the DUs to perform several additional steps. First, they need to convert the phrase into a conjunctive query and perform the search. Then, after obtaining the corresponding documents from the CS, they have to decrypt and screen them to find the ones that contain the phrase they are looking for~(\citealt{zheng_symmetric_2021}). As a consequence, this approach can be very inefficient when dealing with large databases. SE schemes with Phrase Search functionality, on the other hand, enable data users to directly search for phrases~(\citealt{sharma_searchable_2023}).\\
 \\
\textbf{Range search}\\
 \noindent SE schemes allowing range queries enable DUs to search for encrypted data within a specific range of values, meaning that they can perform searches based on an interval rather than just on exact matches. For instance, a data user may want to search for all documents in a database that were created between 2020 and 2022 or search for documents whose owners have more than 30 years old. In fact, the last example mentioned belongs to a specific subcategory of range queries called \textbf{comparison queries}~(\citealt{boneh_conjunctive_2007}). This kind of search query can be very useful to avoid multiple single-keyword searches.\\ 
 \\
\textbf{Occurrence search}\\
\noindent SE schemes with this property allow a data user to issue a query to the cloud server that includes both a search term and a value. This value specifies the minimum number of times that the chosen keyword must appear in a document for it to be retrieved in the search results. This feature can be very helpful to measure the importance of our keyword within each document in the database~(\citealt{handa_searchable_2019}).\\
\\
\textbf{Ranked search}\\
\noindent This feature's main purpose is to enhance the data user's search experience by retrieving only relevant documents that match the query. To achieve this, various scoring functions have been adopted from the information retrieval research community, \citet{handa_searchable_2019}. The most frequently used scoring function is known as TF-IDF, where TF stands for Term Frequency, indicating the significance of that keyword in the document, and IDF stands for Inverted Document Frequency. indicating the significance of the keyword over all documents in the database~(\citealt{pham_survey_2019}). \\

%%%%%%%%%%%%%%%%%%%%%%%%%%%%%%%%%%%%%%%%%%%%%%%%%%%
\subsubsection{Other Functionalities}
There are other functionalities that SE schemes may have and are important for some scenarios. For example, regarding access control functionalities, the ability to authorize and revoke DUs is crucial.\\
\\
\textbf{Authorize and Revoke users}\\
\noindent As the name indicates, this feature in a SE scheme means that the data owner can grant new users search capability as well as revoke this ability. There are not many SE schemes with this functionality, and the ones that exist are inefficient and impractical. This is because enabling this functionality requires compromising security~(\citealt{andola_searchable_2022}).\\
\\
\textbf{Static or Dynamic}\\
\noindent Another important functionality in a SE scheme for real-world applications is the ability to allow dynamic updates. A SE scheme is considered static if  no document is added, deleted, or updated after building the encrypted database.  Therefore, the entire set of data has to be available before data encryption.  
    In contrast, a SE system is said to be dynamic if users can modify, add, or delete documents without compromising the security of the encrypted data or the ability to search and retrieve it. Dynamic SE schemes are more flexible and more suitable for real-world applications. However, static SE schemes are typically more efficient and less complicated to implement than dynamic ones, making them a better option for simple scenarios where the data is relatively stable, and no frequent updates are needed. It is important to notice that, in a dynamic and index-based SE scheme,  the index must be updated efficiently when a change is made to the document collection, and this has to be done without adding any leakage~(\citealt{handa_searchable_2019}).\\ 
\\
 \textbf{Verifiability}\\
 \noindent In most applications, the CS in a SE system is considered semi-trusted, meaning it may retrieve incomplete or incorrect results to the DU~(\citealt{handa_searchable_2019}). Therefore, it is important for a SE scheme to allow the DU to verify the search results retrieved by the CS. Such a SE scheme is designated as a verifiable SE scheme. As would be expected, this kind of scheme would suffer from higher computational overhead on all sites (at DO, at DU, and at the CS) when compared to a SE scheme not allowing verifiability~(\citealt{sharma_searchable_2023}).
    Result verification may include: correctness (if the results retrieved correspond to the query); completeness, if all the relevant documents are retrieved; and freshness if the latest version of the documents is returned to the DU~(\citealt{handa_searchable_2019}). \\
    \\  
\textbf{Delegate}\\
\noindent This functionality allows an authorized user to delegate the search capability to another user. This functionality mainly appears in asymmetric SE schemes, sometimes referred to as Public Key Encryption with Keyword Search. These SE schemes include proxy re-encryption, that is, they have a semi-trusted third-party server (proxy server) that converts a ciphertext created using the DO's public key into a ciphertext that can be decrypted by the delegated user~(\citealt{sharma_searchable_2023}). Although this is not a very common type of SE, there are already several works addressing this feature~(\citealt{noorallahzade_survey_2022}).

\section{Homomorphic Encryption}\label{HomomorphicEncryption}

The main idea behind a HE scheme is to allow computations to be performed over encrypted data without the need to decrypt it first. This concept was introduced by \citet{rivest_data_1978} and was originally called ``privacy homomorphism''. 
A HE scheme is, therefore, any encryption scheme where the encryption function is a homomorphism. Formally  speaking, let $M$ denote the set of plaintexts and $C$ the set of ciphertexts. Let $\odot_{M}$ and $\odot_{C}$ be operations in $M$ and $C$, respectively. An encryption scheme is said to be Homomorphic if for any encryption key $k$, the encryption function $E$ satisfies the property below,
\begin{equation}\label{HomoProp}
\forall m_{1},m_{2} \in M, E(m_{1} \odot_{M} m_{2}) \leftarrow E(m_{1}) \odot_{C} E(m_{2}),
\end{equation}
where $\leftarrow$ means ``can be directly obtained from''. That is, in~\eqref{HomoProp}, $E(m_{1} \odot_{M} m_{2})$ can be directly obtained from $E(m_{1}) \odot_{C} E(m_{2})$ without having to perform any decryption.

HE schemes can be divided into three categories depending on the type and number of operations allowed by the system: Partially Homomorphic Encryption (PHE), Somewhat Homomorphic Encryption (SWHE), and Fully Homomorphic Encryption (FHE).

\subsection{Partially Homomorphic Encryption}
After the publication of Rivest et al.'s work, Cryptography researchers began searching for a homomorphic encryption scheme that allowed computations on encrypted data using more than one operation. However, over the next two decades, most attempts resulted in schemes that only allowed one type of operation, which are the ones known as PHE schemes. In these schemes, the homomorphic property is only satisfied by one operation, such as addition or multiplication, an unlimited number of times.

One example of a PHE scheme is the well-known RSA public key cryptosystem, introduced by \citet{rivest_method_1978}. It only supports the usual product operation and is deterministic, meaning that using the same key to encrypt the same plaintext will always result in the same ciphertext.

Some years later, in \citeyear{goldwasser_probabilistic_1982}, Goldwasser and Micali published the first probabilistic PHE scheme which inspired most of the PHE schemes published in the following decades, such as \citet{benaloh_dense_1994}, which was a generalization of Goldwasser et al.'s scheme, and \citet{naccache_new_1998} which was an improvement on Benaloh's scheme. 

\citet{elgamal_public_1985} introduced a PHE which also allows only the usual product, and, \citet{paillier_public-key_1999} published a PHE which allows only the usual sum. This latter scheme is the most frequently used in SE schemes that use HE, as will be seen in Section~\ref{sec:trends} section.

\subsection{Somewhat Homomorphic Encryption}
In 2005, \citet{boneh_evaluating_2005} proposed the first HE scheme capable of performing two operations: an arbitrary number of additions and just one multiplication, followed by an arbitrary number of additions. These are the kind of HE schemes that are called SWHE, since the homomorphic property is satisfied by more than an operation but a limited number of times.

SWHE schemes were proposed shortly after the first HE scheme. However, they were not as appealing as PHE, due to their limitations. Nonetheless, towards the end of the century, they became much more complete, serving as a stepping stone to achieving a FHE scheme, that is, a HE scheme that allows for more than one operation an unlimited number of times. In fact, many researchers believe that the scheme of \citet{boneh_evaluating_2005} was an important stepping stone towards an FHE scheme.

\subsection{Fully Homomorphic Encryption}
In these type of HE schemes,  the homomorphic property is satisfied by a set of operations an unlimited amount of times.

This category is considered by many as the ``Holy Grail'' of Cryptography, since it allows computations to be performed freely over encrypted data. It was in \citeyear{gentry_fully_2009} that Gentry published the first FHE scheme. In his thesis, Gentry not only proposed an FHE scheme but also provided a method for constructing a general FHE scheme from a scheme with limited but sufficient homomorphic evaluation capacity~(\citealt{marcolla_survey_2022}). Since then, HE has triggered significant interest, and novel constructions on FHE have been proposed following Gentry’s idea, being BGV~(\citealt{brakerski_leveled_2012}), FV~(\citealt{fan_somewhat_2012}), TFHE~(\citealt{chillotti_tfhe_2020}), and CKKS~(\citealt{cheon_homomorphic_2017}) the most representative. 
 
A security liability that is common among HE schemes is the accumulation of encryptions of $0$, i.e., if an attacker is able to accumulate enough encryptions of $0$ then the security of the scheme might be at risk. To overcome this problem, many schemes disguise encryptions of $0$ with what is called \emph{noise}. However, this noise tends to increase with the number of computations performed, which may lead to incorrect decryption in the end. Gentry introduced the main idea of bootstrapping to solve this problem. However, bootstrapping is computationally expensive, and its complex mathematical foundations make Gentry's scheme not practical for real-life scenarios.

According to \citet{acar_survey_2018}, there are four main categories of FHE schemes that have been developed since Gentry's work in 2009:
\begin{itemize}
    \item Ideal Lattices - In this category there are essentially optimizations of Gentry's scheme. Examples of this category include the work of \citet{gentry_implementing_2011} and the work of \citet{scholl_improved_2011}. 
    \item Integers - First appearance in 2010, by \citet{van_dijk_fully_2010}. The main motivation behind these schemes its the simplicity of the concept, however they are not practical, which makes this the least favorite category for researchers;
    \item (Rings) Learning With Error, (R)LWE - It was originated in \citeyear{brakerski_fully_2011}, by Brakerski and Vaikuntanathan. These schemes are based on the LWE problem, which is considered one of the hardest problems to solve in practical time, even for post-quantum algorithms. RLWE is an algebraic variant of LWE, which is more efficient for applications;
    \item $N^{th}$ degree-truncated polynomial ring unit - Like FHE schemes, which are known for their efficiency and use of Multi-Key FHE, these schemes allow computations between data encrypted with distinct keys. An HE scheme of this type was first proposed by \citet{lopez-alt_--fly_2012}.
\end{itemize}

Due to its huge potential for cryptographic applications, HE schemes have already been proposed to solve a wide range of problems in several areas, including in Big Data and Cloud Computing, Secure Image Processing, Medical applications, Electronic Voting Systems, Private Information Retrieval, and Biometric Verifications, as referred by \citet{challa_homomorphic_2020} and \citet{alloghani_systematic_2019}.

%%%%%%%%%%%%%%%%%%%%%%%%%%%%%%%%%%%%%%%%%%%%%%%%%%%%%%%%%%%%%%%%%%%%%%
%%%%%%%%%%%%%%%%%%%%%%%%%%%%%%%%%%%%%%%%%%%%%%%%%%%%%%%%%%%%%%%%%%%%%%
\section{Analysis of SE schemes that utilize HE}\label{sec:analysis}
In this section, we review and analyse the 23 research works which resulted  from the selection process presented in Section~\ref{introduction}. We only considered
the proposed schemes in which  it is clear how HE is used. The goal of this analysis is to study their functionalities, taking in consideration the characteristics presented in Figure~\ref{Figure2}, and identify which of those characteristics leverage the utilization of HE and the type of HE which is used. Table~\ref{tab:categorization} list the selected works and gives an overview of their functionalities. It also identifies the characteristics that are achieved using HE. This table serves as a complete resource for identifying and accessing the works that were analysed in this study, ensuring transparency and facilitating future references.

\begin{table*}[ht]
\caption{Categorization of selected SE schemes that utilize HE}
\label{tab:categorization}
\large
\resizebox{\textwidth}{!}{
\begin{tabular}{ccccccccccccccccccc}
\toprule
\\
\multirow{4}{*}{Article} & \multicolumn{2}{c }{Search Structure} & & \multicolumn{8}{c }{Search Functionalities}    & & \multicolumn{3}{c }{Other Functionalities} & &\multicolumn{2}{c }{User}\\\\  \cline{2-3} \cline{5-12} \cline{14-16} \cline{18-19}
\\
& Index & \begin{tabular}[c]{@{}c@{}}Seq. \\ Scan\end{tabular} && \multicolumn{1}{c }{\begin{tabular}[c]{@{}c@{}}Single \\ Keyword\end{tabular}} & \multicolumn{1}{c }{\begin{tabular}[c]{@{}c@{}}Multi \\ Keyword\end{tabular}} & \multicolumn{1}{c }{\begin{tabular}[c]{@{}c@{}}RE/ \\ Wildcard\end{tabular}} & \multicolumn{1}{c }{Conj.} & \multicolumn{1}{c }{Disj.} & \multicolumn{1}{c }{Range} & \multicolumn{1}{c }{Phrase} & \begin{tabular}[c]{@{}c@{}}Ranked \\ Search\end{tabular} && \multicolumn{1}{c }{Verif.} & \multicolumn{1}{c }{\begin{tabular}[c]{@{}c@{}}Auth./ Revoke\\ User\end{tabular}} & Dynamic & &\multicolumn{1}{c }{Single} & Multi \\ \\\hline
\\
~\citet{malik_homomorphic_2023}  &  & x & &x & &  &    &   &   &   &  &  &  & &&& x & \\\\

~\citet{iqbal_novel_2022} &  & x & &x & &  &    &   &   &  & &  &  &  & && x & \\\\

~\citet{liu_fase_2022}  & x &  &  & & x &  &    &   &   &   & h &  & &  & &&& x \\\\

~\citet{gan_towards_2022}  & x &  & &x & &  &  &  &   &   &   &  &  &  & h && & x \\\\

~\citet{wang_achieving_2022}& x &  &  & &x &  & h  &   &  & & &   &  &  &  &  & & x \\\\

~\citet{andola_secure_2022}   & x &  & & & x &  &    &   &   &   & h & & &  & & & & x \\\\

~\citet{choi_compressed_2021} &  & x && x & &  &    &   &   &   & &  &  &  & & & x & \\\\

%~\citet{abbar_cloud-based_2021}   & x & & & x & &  &    &   &   &   & h & & &  & & & x & \\\\

~\citet{yin_achieve_2021}  & x &  &&  & x & h &  h  & h  &   &   &  &  & & &  & & x & \\\\

~\citet{prakash_pindex_2021}   & x & & &  & x &  &    &   &   &   &  &  & &  & h & & & x \\\\

~\citet{tosun_fsds_2021}   & x &  & & & x &  &    &   &   &   & h & &  &  & & & & x \\\\

~\citet{hou_privacy-preserving_2021}  & x &  & &  & x &  &  x  &   &   & h  &  &  & & & h & & & x\\\\

~\citet{zhang_secure_2020}  & x &  && x & &  &    &   &   &   & h & &  &  & & & x & \\\\

~\citet{elizabeth_verifiable_2020}  & x &  &&  & x &  &    &   &   &   & h & & x &  & h && & x \\\\

~\citet{li_efficient_2020}   & x &  & & & x &  &    &   &   &   &  h &  & & & h & & x & \\\\

~\citet{yang_flexible_2020}   & x &  & & & x & h &  h  & h  &   &   & h &  & & x & & & & x \\\\

~\citet{wen_leaf_2020} &  & x && x & &  &    &   &   &   &  &  & &  & & & x & \\\\

~\citet{yang_multi-user_2020}   & x &  &&  & x &  &    &   &   &   & h &  & & x & & & & x \\\\

~\citet{boucenna_secure_2019}  & x & & &  & x &  &    &   &   &   & h & & &  & & & & x \\\\

~\citet{guo_secure_2019} & x &  &  &&  &  &    &   &  h &   &  &  &  & & & & & x \\\\

~\citet{shen_secure_2019}    & x &  &&  & x &  &  x  &   &   & h  &  &  & & & & & & x \\\\

~\citet{akavia_secure_2018}  &  & x & &x & &  &    &   &   &   &  &  &  & & & & x & \\\\

~\citet{elizabeth_tsed_2018}  & x &  &  & &x &  &    &   &   &   & h &  & & & h & & & x \\\\

~\citet{wu_verifiable_2018} & x &  & & & x &  & h   &   &   &   &  && h & & & & & x \\\\

\bottomrule \\
\multicolumn{6}{l}{x: indicates that a scheme has a certain property}\\
\multicolumn{6}{l}{h: indicates that a scheme uses HE to achieve a certain property}
\end{tabular}
}

\end{table*}

Our analysis is divided into the following categories: Search Structure, Search Functionalities, and Other Functionalities. We have not devoted a section to the category ``Multiplicity of users'' because this functionality is mentioned whenever we analyse one of the selected works. It is important to notice that it is out of the scope of this work to provide a detailed explanation of the cryptographic constructions, which can be found in the referenced works.

\subsection{Search Structure}
The search structure  of a SE scheme can involve either a sequential scan of the entire database or an index-based approach, where an index is created to facilitate the search process, as mentioned in Section~\ref{SearchableEncryption}. In this analysis, we will focus on the sequential scan-based SE schemes, as all the others are index-based, and they will be discussed in the next sections. Additionally, it is important to recall  that the concept of secure search is similar to that of SE, and consequently, both approaches will be included in this analysis. Notice that secure search methods typically involve verifying all the documents in the database to find the ones that match the query, which is similar to a sequential scan process. 

The work of \citet{akavia_secure_2018} is the first of the selected papers to introduce a method to perform secure search on FHE encrypted data. The authors claim that this is realized by a polynomial of degree logarithmic (rather than linear) in the number of array entries data. The core of their secure search protocol is the computation of a sketch of the first query match, which is done using an approach they call SPiRiT. This method returns the first record in an encrypted array of data matching a lookup value. Their proposed scheme can use any standard semantically secure FHE in which  a prime number $p$ can be chosen as the plaintext modulus parameter so that the homomorphic operations are addition and multiplication modulo $p$. They claim to have a single communication round and that the communication overhead only grows with input and output sizes.

In \citeyear{wen_leaf_2020}, Wen et al. proposed a searching method to be used in a secure search scheme, meant for a single-user setting, named LEAF. This method relies on three novel techniques: Localization, Extraction, and Reconstruction. Localization is used to divide the original database array into smaller intervals of equal length and to identify the first one that contains a matching item. It returns the encrypted indexes of the interval containing the matched item. Extraction is used to extract the interval containing the first matched item for subsequent search operations on that interval and Reconstruction is used to combine the information from these two techniques in order to properly locate our desired data, without needing to decrypt anything. It is worth mentioning that these authors claim that secure search roughly contains two steps, i.e. \emph{matching}, and \emph{searching}. In the matching step, the server compares the encrypted search query (from the client) with all encrypted items in the database, and returns another encrypted array of 0s and 1s with 1 indicating the corresponding database item satisfying the query. The searching step returns all 1's indexes and corresponding items to the client. Therefore, their work focuses on the searching step\\
FHE is used in this scheme to encrypt both sensitive data and search queries. In fact, one of the main goals of this scheme is to optimize the use of FHE in Secure Search by reducing the number of necessary computations, namely the number of multiplications.\\
Moreover, the authors also propose a variant of this scheme, named LEAF+, which uses lazy bootstrapping. On the upside, the bootstrapping step can control the computational depth required by the algorithm, and the greater the database size, the greater the optimization effect. On the other hand, when the size of the database is small, this variant will be ineffective, since it brings many extra multiplication operations and computation depth.

Choi et al., in \citeyear{choi_compressed_2021}, suggested a secure search method that uses a standard CPA-secure (leveled)  FHE, meant for single use, that also performs a sequential scan over the encrypted database to perform a search. In fact, in their approach, the computational task of secure search is divided in two steps called \textit{matching} and \textit{fetching}, which correspond to the \emph{mathching} and \emph{searching} phases of Wen et al.'s approach. In the matching step, the cloud server compares the encrypted search query with all encrypted records in the database using the homomorphic properties of the underlying FHE scheme to find the ones that correspond to that query. Then, in the fetching step, those corresponding records are retrieved from the database, also using the homomorphic properties, and made available to the data user, which can then decrypt them. Regarding the fetching procedure, the authors propose two novel retrieving algorithms, namely the COIE scheme (based on power sums or bloom filters) and CODE scheme (based on bloom filter sets), and compare the performance of these algorithms with other well known retrieving algorithms like LEAF+~(\citealt{wen_leaf_2020}) and PIR. 

In \citeyear{iqbal_novel_2022}, Iqbal et al. proposed a mechanism to securely search encrypted audio data that has been outsourced to CS in a medical context, and which also performs a sequential scan. Their approach involves using the BGV FHE scheme to encrypt the data files, which are then sent to the cloud server for storage. When a search is requested, the CS performs a sequential scan on the stored encrypted files using homomorphic operations to find the documents that contain the searched keyword. Then, the retrieved documents are decrypted using the BGV scheme. In the experiments performed by Iqbal et al., the open-source programming library HElib version 2.1.0 was used. This library allows using BGV with bootstrapping and to apply enhancements such as the ciphertext packing technique by Smart–Vercauteren and the optimization technique by Gentry–Halevi–Smart~(\citealt{haleviAlgorithmsHElib2014}).

In the recent work of Malik et al., published in \citeyear{malik_homomorphic_2023},  a Single Keyword SE scheme that uses PHE is presented, which uses the Paillier cryptosystem~(\citealt{paillier_public-key_1999}) to protect airport data that is outsourced to cloud servers. This scheme uses HE to perform search operations and provides a high level of security by hiding search patterns using trapdoors. Two approaches were proposed, one which exploits the deterministic properties of the Paillier cryptosystem, referred to as the ``efficient SKSE'', and the other which takes advantage of its probabilistic properties, named the ``secure SKSE''. The former is suitable for scenarios that prioritize a lightweight approach over security. In contrast, the latter is more appropriate for scenarios in which the security has priority over performance. Both approaches use a sequential scan search structure and are designed for a single-user scenario where the airport acts as both the DO and DU of the encrypted data. 

In this scheme, original data files are encrypted twice. The first encryption uses the Advanced Encryption Standard (AES)~(\citealt{117146}), after which the encrypted files are uploaded to the CS.  The second encryption uses the Paillier cryptosystem to encrypt the original files, and the resulting encrypted data is also uploaded to the CS. These are the encrypted files that will be used to perform the sequential scan. As an output of this scan, the system retrieves an encrypted result which, after being decrypted by the DU using Paillier's scheme, allows recovering the identifiers of the files containing the searched keyword. AES is then employed to decrypt these files and recover the original data.

\subsection{Search Functionalities}
In this section, we analyze the selected works that address search functionalities. We have divided this section based on the following characteristics: Regular Expressions and Wildcards, Conjunctive Search, Range Search, Phrase Search, and Ranked Search. We do not provide separate sections devoted to the ability to allow a multi-keyword search or disjunctive search, because all the works that possess those functionalities also possess other search functionalities and are analysed in their corresponding sections. Also, the papers that have a specified characteristic can be easily identified in Table~\ref{tab:categorization}.
%%%%%%%%%%%
\subsubsection{Regular Expressions and Wildcards}
SE Schemes allowing wildcard queries are spare, and only two were found during our research. The first scheme in question was proposed by Yang et al. in \citeyear{yang_flexible_2020}. In that work, they presented a novel index-based SE scheme designed for a multi-user environment. The proposed scheme allows wildcard queries as well as user authorization and revocation. Additionally, the user can issue “AND” or “OR” queries on search keywords and can also obtain the top-$k$ documents that have the highest relevance scores.

This scheme uses a simple index approach. More specifically, a document index contains three pieces of information: the document ID, its corresponding keywords, and the key utilized in the symmetric encryption scheme used to protect the documents. The index is encrypted using a PHE scheme, namely Paillier's cryptosystem, before being outsourced to the cloud server. On the other hand, the documents themselves are encrypted with any secure symmetric encryption scheme.

The authors cover a wide range of different algorithms to look for a match within the search protocols, regarding different types of wildcard queries. These algorithms can be split into three categories: zero wildcards (1 algorithm), one wildcard (3 algorithms) and two wildcards (4 algorithms). More specifically, for  one wildcard we have the following possibilities for a query: $\text{*}+ Y_1$, $Y_1 + \text{*}$ and $Y_1 + \text{*}+ Y_2$, ($Y_1, Y_2$ are strings of any size). When two wildcards are present in the query, we have the following possibilities: $\text{*}+ Y_1 + \text{*}$, $Y_1 + \text{*} + Y_2 + \text{*}$, $\text{*} + Y_1 + \text{*}+ Y_2$ and $Y_1 + \text{*}+ Y_2 + \text{*} + Y_3$ ($Y_1, Y_2, Y_3$ are strings of any size). These algorithms take advantage of the homomorphic properties of the Paillier's cryptosystem to encrypt the keywords in a way that  comparisons are possible to identify matches. Finally, with respect to the authorization and revocation functionality, PHE is not utilized. Despite this, the system allows a DO to grant research privileges to other users for a specified period of time and to automatically revoke these privileges once the authorization period expires.

In \citeyear{yin_achieve_2021}, Yin et al. suggested a new SE scheme which uses FHE and allow some types of wildcard queries. The scheme was designed to achieve compound substring query on multiple attributes. A substring query, as the name suggests, is a query that allows to search for a contiguous sequence of characters within a string. For example, the substring query ``cat'' would return all the results which contain the substring cat, e.g. caterpillar, cats, concatenate, ducat. The proposed scheme can, in fact, support two types of substring patterns: $\text{*} + s + \text{*}$ and $s1 + \text{*} + s2$, where $s$, $s1$, and $s2$ represent queried substrings and $\text{*}$ represents any string of any length. This is achieved by constructing a tree index structure using a modified version of the well-known position heap technique.

In this scheme, the sensitive data is encrypted using a symmetric key encryption scheme indistinguishable under a chosen-plaintext attack, such as AES, and the tree index is encrypted using an FHE scheme, such as BGV, and a pseudorandom function. The FHE scheme is used to encrypt the ID of each node in the tree, and the pseudorandom function is used to encrypt the concatenation of all edges labels along the path from the root to that node.  

Based on the properties of the FHE scheme, the authors also designed an algorithm to calculate the intersection of search results for different searched keywords and therefore achieve the compound substring query on multiple attributes. The compound formula, in this case, consists of conjunctive and disjunctive expressions of the substrings queried and, consequently, this scheme allows conjunctive and disjunctive queries as defined in Section~\ref{SearchableEncryption}.

We have not found any literature proposing a SE  scheme that enables the use of regular expressions and utilizes HE.
%%%%%%%%%%%%
\subsubsection{Conjunctive Search}
The ability to perform a conjunctive search is very attractive, and among the selected works, six of them allow this functionality. However, only 4 of those papers make use of HE to facilitate the conjunctive search. Most of those papers possess other search functionalities and are analysed in the corresponding sections. Interestingly, only one work provides conjunctive search and no other search functionality. 
This work is duo to Wang et al., which, in \citeyear{wang_achieving_2022}, proposed an index-based SE scheme, for a multi-user setting, that supports conjunctive keyword search and uses a special PHE scheme to hide the search pattern. Specifically, their work exploits the additive homomorphic encryption scheme PBC~(\citealt{bressonSimplePublicKeyCryptosystem2003}), along with an auxiliary server, to efficiently achieve the conjunctive keyword search property while ensuring that the data user cannot learn anything other than the desired search result. In fact, the auxiliary server is introduced to allow the system to achieve the desired properties by adopting what the authors refer to as the double trapdoor decryption mechanism, which is available in this scheme. 

 To achieve the conjunctive keyword search property, the authors use the polynomial representation of a multiset and encrypt the polynomials using the BCP cryptosystem to maintain their confidentiality. The cloud server and the auxiliary server then work together using both additive and multiplicative homomorphic properties to perform the search.

 This work does not detail what kind of encryption scheme is used to encrypt the documents before uploading them to the database.

%%%%%%%%%%%
\subsubsection{Phrase Search}

The use of Homomorphic Encryption to enable Phrase Search in SE schemes is not very common. In fact, only 2 out of the 23 analysed SE schemes are capable of performing phrase search and use HE to allow that functionality. The first one was proposed in \citeyear{shen_secure_2019}, by Shen et al, and is named P3. This scheme is meant for a multi-user setting and supports multi-keyword search, specifically phrase search, and conjunctive keyword search.
The main idea of the scheme is to build an inverted index containing not only the document identifiers where each keyword appears, but also the location of each keyword in those documents. 

To protect the location information, the scheme encrypts it using a PHE scheme from \citet{boneh_evaluating_2005}, whose homomorphic properties allow the server to analyse if two encrypted keywords are adjacent.  Moreover, this enables the DU to obtain precise search results from a single interaction with the cloud server.
Note that phrase search is a specific type of conjunctive query where the position of queried keywords matters.

The document identifiers and keywords are protected using other techniques, specifically a pseudorandom permutation primitive and a secure kNN technique, respectively.

Subsequently, in \citeyear{hou_privacy-preserving_2021}, Hou et al. proposed a SE scheme that also uses the PHE scheme from Boneh et al. to protect the location of keywords and enable phrase search. However, they propose a different structure for the index, which they call a virtual binary tree. This index tree is only a logical structure used to store the keywords and related information. Its elements are stored in a hash table, if they are a leaf node, or are mapped to a bloom filter otherwise. Homomorphic properties are used similarly to the work of \citet{shen_secure_2019}, allowing the scheme to check if the keywords are adjacent. Additionally, their approach allows for dynamic updates, which is an advantage when compared to the previous scheme.

%%%%%%%%%%%
\subsubsection{Range Search}

The ability to perform range search is another functionality which is not usual to have in SE schemes that use HE. In our research, only one work, duo to Guo et al., was found. This work was published in \citeyear{guo_secure_2019}, and it presents a probabilistic threshold range search scheme meant for the multi-user setting in the IoT context. This scheme addresses the problem of performing range searches over multidimensional uncertain data while allowing false positives by returning all the encrypted data that have a probability, of being within the range of interest, higher than a given threshold value. 

The concept of the scheme is that DOs obtain uncertain data from the IoT devices, which are modelled as multidimensional objects. As such, each IoT object is represented by an uncertain region and its probabilistic density function. A piece of data collected from such an object is called instance and comprises three components: the identification of the object, the coordinates of the instance, and its probability.

Their approach utilizes a KD-tree, which is a data structure for indexing $d$-dimensional point data distributed in a $d$-dimensional space. Each node in the KD-tree consists of one instance and its corresponding range, calculated based on the instance's coordinates. The nodes are encrypted using an Order Preserving Encryption (OPE) scheme, while the instance probabilities are encrypted using Paillier's cryptosystem

It is noteworthy that the scheme relies on two cloud servers. Specifically, the first cloud server is responsible for: storing both the encrypted data and the encrypted KD-tree, performing KD-tree searches, and sending the results to the second server. Then, the second server is entitled to filter the results based on the threshold provided in the search query, and sending the obtained encrypted documents to the DU.

When the first cloud server receives a query, it conducts a search over the KD-tree by comparing the values of the instances' coordinates and using the homomorphic addition of the Paillier cryptosystem to calculate the upper and lower appearance probability of each IoT object with respect to the search, so the results fall within the range queried. The results are then sent to the second server for filtering, and finally, the filtered results are sent to the data user.

%%%%%%%%%%
\subsubsection{Ranked Search}\label{sec:ranked}
The ability to perform a ranked search is the most frequently observed feature in SE that use HE, with 10 out of the 23 analysed works having this property. Furthermore, all of these works utilize HE to achieve that functionality.

In \citeyear{elizabeth_tsed_2018}, Elizabeth et al. proposed a multi-user Top-$k$ Ranked SE (TSED) scheme that allows dynamic updates.  The scheme is based on an inverted index, which uses a binary vector for each keyword to indicate whether each file contains that keyword (similar to the Z-index proposed by Wu et al. in~\citeyear{wu_verifiable_2018}).

In this scheme, sensitive data is encrypted using a secure symmetric encryption scheme, such as AES. Two PHE schemes are then used to encrypt the two components of the inverted index. Specifically, Paillier's cryptosystem is used to encrypt each keyword, while Goldwasser-Micali (GM)~(\citealt{goldwasser_probabilistic_1982}) is used to encrypt each binary index vector.  

TSED scheme allows for top-$k$ ranked searches using the TF-IDF rule. This process is done by a secure coprocessor which is responsible for computing the scores for the query keyword using the encrypted score index, ranking according to its relevance to the query, and returning the top-$k$ document identifiers to the cloud server which then retrieves the corresponding documents do the data user.

A variation of this scheme was proposed in \citeyear{elizabeth_verifiable_2020}, also by Elizabeth et al., which was designated verifiable top-$k$ ranked SE over encrypted cloud data with dynamic updates~(VSED). This scheme, besides its dynamic and ranked search capabilities similar to the VSED scheme, it also has the verifiability capability.

 In VSED scheme, a similar encrypted inverted index is constructed, but now the authors use a secret orthogonal vector and the Paillier cryptosystem~(\citealt{paillier_public-key_1999}). Both the index and a trapdoor which is used when performing searches are encrypted using the PHE scheme, as well as the ranking score which is computed beforehand and, similarly to what is done in TSDE, a secure coprocessor is used to find the top-$k$ ranked searches  based in the TF-IDF rule. 
 
The Paillier cryptosystem is also used in the update process, which is done using two different algorithms: one to add new keywords and the other one to delete existing ones.

To verify the query results returned by the cloud server, this scheme uses the well known mechanism for message authentication named HMAC~(\citealt{rfc2104}).

Regarding the sensitive data, it can be encrypted using a secure symmetric encryption scheme such as AES, as in TSED.

In \citeyear{boucenna_secure_2019}, Boucenna et al. proposed a SE scheme, named Secure Inverted Index based SE (SIIS), meant for a multi-user setting, which supports ranked search using HE. In fact, this scheme also allows user access rights management by using a CP-ABE to encrypt the data collection. As the name of the scheme indicates, it involves the construction of inverted indices. More specifically, two separate inverted indices are used. One is used for storing similarity scores, which are computed by applying the double score weighting formula~(\citealt{boucennaConceptbasedSemanticSearch2016}). The second inverted index is constructed in order to manage the users’ access rights to the data collection, and reduce the number of false positives generated by using the dummy documents technique~(\citealt{caoPrivacyPreservingMultiKeywordRanked2014}) to allow keyword privacy. The entries in this second index correspond to the users' IDs, which points to the set of documents to which they have access. Both indexes are then encrypted using the FHE scheme BGV whose homomorphic properties are used to perform the search.

In \citeyear{zhang_secure_2020}, Zhang et al. proposed a secure ranked search model over encrypted data in hybrid cloud computing, meant for a single-user setting. Hybrid cloud computing refers to the use of a public and private cloud, where the latter is mainly used to perform costly computations that would have to be performed on the client's side otherwise. In this scheme, the private cloud has the ability to encrypt and decrypt data and most communication rounds are done between the private and public clouds.

In this proposal, the authors use the Ocapi BM25 ranking model~(\citealt{whissellImprovingDocumentClustering2011}), and a not specified FHE scheme is used to encrypt the term frequency TF and inverse document frequency IDF separately, on the private cloud. This information is then used to build an inverted index, which is done in the public cloud. 

Regarding the encryption of the sensitive data, it can be performed with any secure symmetric encryption scheme, such as AES. Finally, the proposed scheme also relies on a retrieval technique to be used by the private cloud to download the encrypted documents, although none is specified by the authors.

In the same year, \citet{li_efficient_2020}, proposed a Two-Server Ranked Dynamic SE scheme (TS-RDSE), meant for a single-user setting, which supports multi-keyword search. This scheme uses two cloud servers to perform the searching and sorting, with one of them being responsible for storing the encrypted data and the other one for storing the secret key. This scheme uses orthogonal vectors and two PHE schemes, namely Paillier's and GM, in the  encryption of an inverted-index, which contains TF-IDF scores that are used to perform ranked search. More specifically, the index is divided into a search index, which uses both Paillier and GM for encryption, and a weight index, which uses only Paillier.

Since the encryption of the index is based on both PHE schemes, TS-RDSE protocols for adding or deleting documents also take advantage of the homomorphic properties in order to efficiently update the index. 

Regarding the encryption of sensitive data, it can be encrypted with any secure symmetric encryption scheme.

In \citeyear{yang_multi-user_2020}, Yang et al. proposed an index-based SE scheme meant for the multi-user setting which support multiple data owners. This scheme also supports multi-keyword queries, top-$k$ ranked results and user authorization/ revocation.

The authors use a PHE scheme, more specifically the Paillier cryptosystem with threshold decryption (PCTD) to encrypt the index, which is constituted by the keywords along with their weight, documents IDs and documents keys (the weight of a keyword can be computed using TF-IDF rule, for example). Given a certain query, the authors propose a novel algorithm to compute the relevance scores regarding such query, named Secure Multiple Keyword Search Protocol Across Domains (MKS).

The encryption of the files themselves is performed by any secure symmetric encryption scheme.

 In \citeyear{tosun_fsds_2021}, Tosun et al. proposed a new multi-user SE scheme, named Fully Secure Document Similarity~(FSDS) which combines Secure K Nearest Neighbour~(SK-NN), a secure algorithm that operates on data sets in Euclidean space and measures similarity using Euclidean distance, with SWHE.  In this scheme, the sensitive data and the search queries are represented as TF - IDF vectors which are encrypted with a variant of SK-NN that the authors named mSk-NN.  The searchable index, which is generated from the documents using TF-IDF representation, is encrypted first with the mSK-NN, and then by a SWHE scheme named FV~(\citealt{fan_somewhat_2012}). The overall concept of the scheme is to compute the $k$ nearest neighbours to a given query using the cosine similarity comparison metric. The authors claim that using a combination of mSK-NN and SWHE results in a more efficient and secure system, compared to an approach that only uses SWHE to encrypt the queries. This is because the amount of computations required is minimized.

Liu et. al proposed, in \citeyear{liu_fase_2022}, an index-based multi-keyword SE scheme which uses FHE to support ranked search, namely, to retrieve the top-$k$ documents most relevant according to the search query. The documents themselves are encrypted using a symmetric encryption scheme (not specified), while an FHE scheme is used to encrypt the indexes and the search queries. The FHE used was developed by the same authors and is named  Fully Homomorphic Order-Preserving Encryption~(FHOPE)~(\citealt{liu_novel_2018}). This encryption method supports homomorphic addition, homomorphic multiplication, and order comparison over encrypted data. As such, it is used to support the search operation and calculation of relevance score over the encrypted data. In this system, the data owner is responsible for extracting the keywords, encrypting the searchable index with FHOPE and uploading it to a cloud server. The data user is responsible for sending the search query to the cloud server, which performs the search operation and the ranking score operation, and it then returns the most relevant top-$k$ documents to the data user.

In the same year, \citet{andola_secure_2022} proposed a SE scheme, meant for a multi-user setting, that supports multi-keyword queries and uses HE properties to construct and search over an index. This scheme also allows ranked search, which is achieved using term frequency and inverse document frequency TF-IDF.  The authors use Elliptic Curve based ElGamal~(\citealt{menezes_1996}) to encrypt the index, and regarding the encryption of sensitive data, it can be performed with any secure symmetric encryption scheme.

%%%%%%%%%%%%%%%%%%%%%%%%%%%%%%%

%%%%%%%%%%%%%%%%%%%%%%%%

\subsection{Other Functionalities}
In this section, we analyse the selected works that offer other functionalities, although they are not common. Specifically, we did not  find any papers that use HE to provide the ability to ``Authorize or Revoke'' access, nor did we find any works addressing the``Delegate'' functionality. As for the remaining functionalities, we have identified 6 approaches that are ``Dynamic'' and just one which uses HE to provide the ``Verifiability'' characteristic.

\subsubsection{Dynamic}

Even though dynamic SE schemes are more likely to be used in real world applications,  not many works have been proposed in the literature to address this problem. In our research, we identified only five published studies that exploit the HE properties to develop dynamic SE schemes. There are, however, other schemes, which are dynamic, but they do not utilize the HE properties to achieve this functionality, as can be seen in Table~\ref{tab:categorization}.

Before 2021, three dynamic SE schemes that use HE to provide index updates were proposed, namely the TSED scheme proposed by \citet{elizabeth_tsed_2018}, a variant of this scheme, named VSED~(\citealt{elizabeth_verifiable_2020}), and the TS-RDSE scheme proposed by \citet{li_efficient_2020}. These schemes are designed for a single-user setting.  

Prakash et al. proposed, in \citeyear{prakash_pindex_2021}, a dynamic and index-based SE scheme, named PINDEX, which is intended for a multi-user setting. Their approach suggests a dynamic index construction method that is multi-linked, and which uses the PHE scheme proposed by Paillier, to encrypt the index, along with secret orthogonal vectors as building blocks.  In this scheme, a DO can add and delete keywords or documents without reconstructing the outsourced encrypted index, and the homomorphic properties are used on both the search and update processes.

Furthermore, the proposed scheme achieves forward privacy due to the probabilistic nature of the Paillier cryptosystem and the use of secret orthogonal vectors.  Notice that, in a dynamic SE scheme, forward privacy is a critical requirement since it ensures that newly added data does not reveal any information about previously searched queries. This is especially important in a multi-client setting where different users search different queries, and new data should not reveal anything about previous searches.

 In \citeyear{gan_towards_2022}, Gan et al. proposed a new dynamic searchable symmetric encryption scheme for multi-client settings that uses XOR homomorphic function to ensure forward privacy. To achieve efficient multi-client search and forward privacy, the authors introduced two novel data structures: private links and a public search tree. During the search process, each client can choose to search their own private link or the public search tree, or both.  The proposed scheme also employs a state-based approach to manage database updates. This involves maintaining a state variable that tracks the current state of the database and allows for efficient updates without requiring its complete re-encryption. The XOR-homomorphic function is used in both the update and search processes. The documents themselves are encrypted using a symmetric cryptosystem.

\subsubsection{Verifiability}

There is only one paper which uses HE to provide the verifiability characteristic. In fact, only one more paper was found in our research which has this property, but does not use HE to achieve it, namely the work of \citet{elizabeth_verifiable_2020}.

The former paper is due to Wu et al. and it was published in \citeyear{wu_verifiable_2018}. Their work proposed a Verifiable Public Key Encryption With Keyword Search based on HE meant for the multi-user setting.\\
In this scheme, sensitive data is encrypted using a standard proxy re-encryption public key algorithm. The authors also suggest a novel index construction, named Z-index, which uses a binary vector for each keyword to indicate whether each file contains that keyword. The index is constructed using an inverted data structure and is encrypted using an  FHE scheme, namely DGHV~(\citealt{van_dijk_fully_2010}), and a homomorphic hash function. The main purpose of this scheme is to avoid the use of query trapdoors and therefore improve search efficiency. The verifiability feature is achieved using the homomorphic hash function.

\section{Research Trends and Discussion}\label{sec:trends}

The research on SE schemes enhanced by HE properties has significantly increased in recent years. This trend aligns with the broader adoption of HE in various domains.  In this section, we aim to provide an overview of the current trends in this research area and highlight their significant implications on the design of the studied approaches. Specifically, we will discuss the types of HE shcemes used in SE, the application of HE in search structures, in search functionalities, and in other types of functionalities. Finally, we will also discuss the ability to allow multi-users, even though achieving this functionality does not directly rely on HE.

\subsection{Types of HE schemes used in SE}
The use of HE in SE schemes has increased in recent years. In fact, 74\% of the 23 selected works were published within the last three years. This is consistent with the overall trend observed in the  increased application of HE in diverse areas.

\begin{figure}[ht]
\centering
\includegraphics[width=0.8\columnwidth]{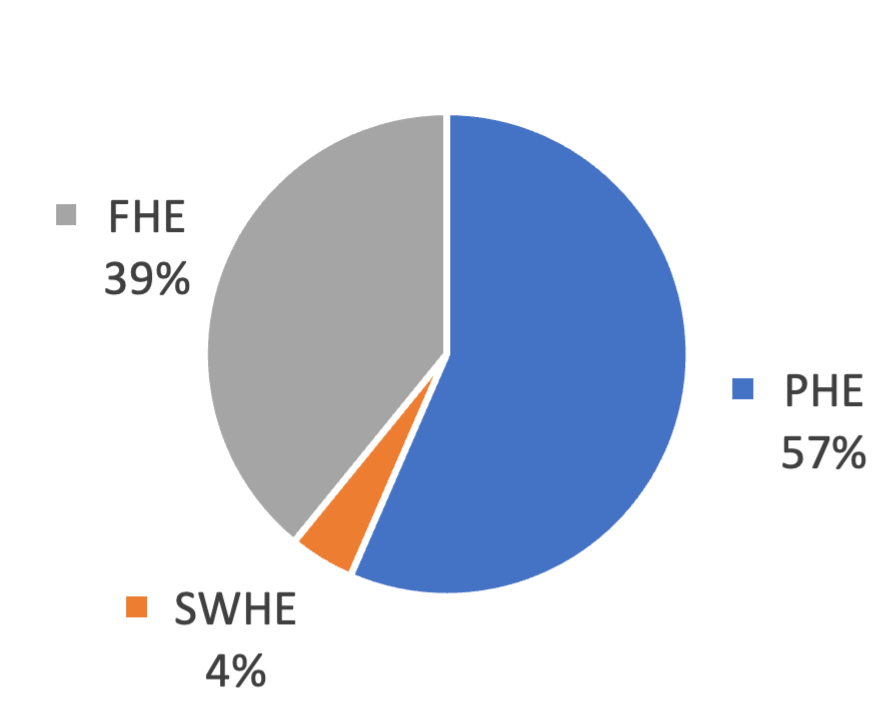}
\caption{Types of HE schemes used in SE}
\label{Figure3}
\end{figure}

Our analysis revealed that PHE is the most frequently type of HE  used in SE, as shown in Figure~\ref{Figure3}. This is not surprising since PHE is the simplest type of HE among the three. On the other hand, FHE schemes have the most potential, but they are still inefficient. However, despite this, a considerable percentage of SE schemes  utilize FHE due to its potential benefits (38\% use FHE for 57\% of systems using PHE).

\begin{table*}[ht]
\caption{HE schemes used in selected SE approaches}
\label{tab:typeofHE}
%\large
\resizebox{\textwidth}{!}{
\begin{tabular}{ccccccccccccccc}
\toprule
\\
\multirow{4}{*}{Article} & \multicolumn{7}{c}{PHE} & & SWHE & & \multicolumn{4}{c}{FHE} \\\\ \cline{2-8} \cline{10-10} \cline{12-15} \\
& Paillier & EC El Gamal & XOR & Boneh & PCTD & GM & BCP & & FV & & BGV & FHOPE &  BFV  & DGHV \\\\ \hline
\\
~\citet{malik_homomorphic_2023} & x &   &   &  & & & & & & & & & & \\\\

~\citet{iqbal_novel_2022} &  &   &   &  & & & & & & & x & & & \\\\

~\citet{liu_fase_2022} &  &   &   &  & & & & & & & & x & & \\\\

~\citet{gan_towards_2022} &  &   & x &  & & & & & & & & & & \\\\

~\citet{wang_achieving_2022} &  &   &   &  & & & x &  & & & & & & \\\\

~\citet{andola_secure_2022} &  & x  &   &  & & & & & & & & & & \\\\

~\citet{choi_compressed_2021} &  &   &   &  & & & & & & & & & & \\\\

%~\citet{abbar_cloud-based_2021} &  &   &   &  & & & & & & & & & x & \\\\

~\citet{yin_achieve_2021} &  &   &   &  & & & & & & & x & & & \\\\

~\citet{prakash_pindex_2021} & x &   &   &  & & & & & & & & & & \\\\

~\citet{tosun_fsds_2021} &  &   &   &  & & & &  & x & & & & & \\\\

~\citet{hou_privacy-preserving_2021} &  &   &   & x & & & & & & & & & & \\\\

~\citet{zhang_secure_2020} &  &   &   &  & & & & & & & & & & \\\\

~\citet{elizabeth_verifiable_2020} & x &   &   &  & & x & & & & & & & & \\\\

~\citet{li_efficient_2020} & x &   &   &  & & & & & & & & & & \\\\

~\citet{yang_flexible_2020} & x &   &   &  & & & & & & & & & & \\\\

~\citet{wen_leaf_2020} &  &   &   &  & & & & & & & & & x & \\\\

~\citet{yang_multi-user_2020} &  &   &   &  & x & & & & & & & & & \\\\

~\citet{boucenna_secure_2019} &  &   &   &  & & & & & & & x & & & \\\\

~\citet{guo_secure_2019} & x &   &   &  & & & & & & & & & & \\\\

~\citet{shen_secure_2019} &  &   &   & x & & & & & & & & & & \\\\

~\citet{akavia_secure_2018} &  &   &   &  & & & & & & & x & & & \\\\ 

~\citet{elizabeth_tsed_2018} & x &   &   &  & & x &  & & & & & & & \\\\

~\citet{wu_verifiable_2018}  &  &   &   &  & & & & & & & & & & x \\\\ \bottomrule

\end{tabular}
}
\end{table*}

Finally, we also observed that more than a half of the schemes that rely on a PHE scheme use either the Paillier cryptosystem (Malik et al.~\citeyear{malik_homomorphic_2023}, PINDEX~\citeyear{prakash_pindex_2021}, Li et al.~\citeyear{li_efficient_2020}, Yang et al.~\citeyear{yang_flexible_2020}, Guo et al.~\citeyear{guo_secure_2019}), a combination of the Paillier cryptosystem with GM (VSED~\citeyear{elizabeth_verifiable_2020}, TSED~\citeyear{elizabeth_tsed_2018}) or the Paillier cryptosystem with threshold decryption (Yang et al.~\citeyear{yang_multi-user_2020}). The reason that so many approaches choose to use the Paillier cryptosystem (or any other variations) is the simplicity and security of the scheme, as well as the number of existing implementations freely available.\\
The remaining articles using PHE, use several different schemes, such as XOR-homomorphic (Gan et al.~\citeyear{gan_towards_2022}), BCP (Wang et al.~\citeyear{wang_achieving_2022}), EC El Gamal (Andola et al.~\citeyear{andola_secure_2022}) and Boneh (Hou et al.~\citeyear{hou_privacy-preserving_2021}, Shen et al.~\citeyear{shen_secure_2019}).\\
SWHE schemes were barely present in the studied schemes. In fact, only one article uses a SWHE scheme, more specifically an implementation of the FV scheme (FSDS~\citeyear{tosun_fsds_2021}). On the other hand, FHE schemes are used in 9 articles, where the BGV (Iqbal et al.~\citeyear{iqbal_novel_2022}, Yin et al.~\citeyear{yin_achieve_2021}, Boucenna et al.~\citeyear{boucenna_secure_2019}, Akavia et al.~\citeyear{akavia_secure_2018}) and BFV (LEAF~\citeyear{wen_leaf_2020}) are utilized the most, leveraging HElib, which is an open-source software library that implements some HE schemes. Table~\ref{tab:typeofHE} lists the selected works and identifies the HE schemes they use.

%%%%%%%%%%%%%%%%%%%%%%%%
\subsection{HE usage in Search Structures}
\begin{figure}[ht]
\centering
\includegraphics[width=0.8\columnwidth]{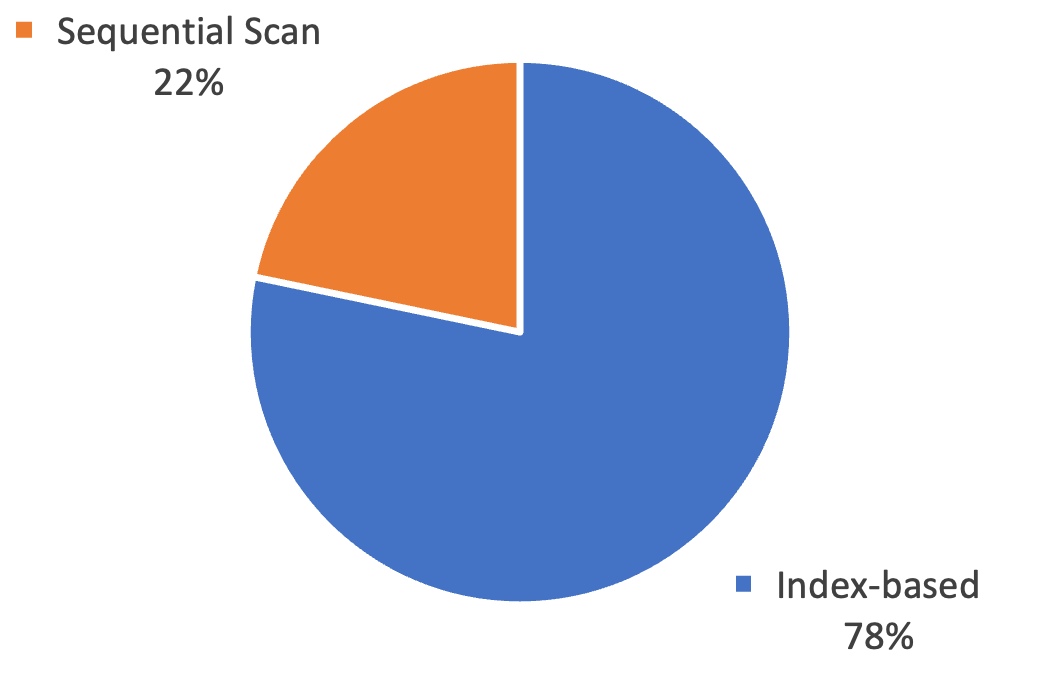}
\caption{Use of HE on the Search structure}
\label{Figure4}
\end{figure}

The utilization of HE in SE schemes has been well explored to provide more efficient search structures. This applies to both sequential scan-based approaches and index-based ones. In all the papers that have been studied, HE is used to design the search structure of the respective scheme. It is worth mentioning that index-based SE schemes are predominant (see Figure~\ref{Figure4}), which is expected since most current applications require efficient search processes on large encrypted databases.

%%%%%%%%%%%%%%%%%%%%%%%%%
\subsection{HE usage in Search Functionalities}
\begin{figure}[ht]
\centering
\includegraphics[width=0.9\columnwidth]{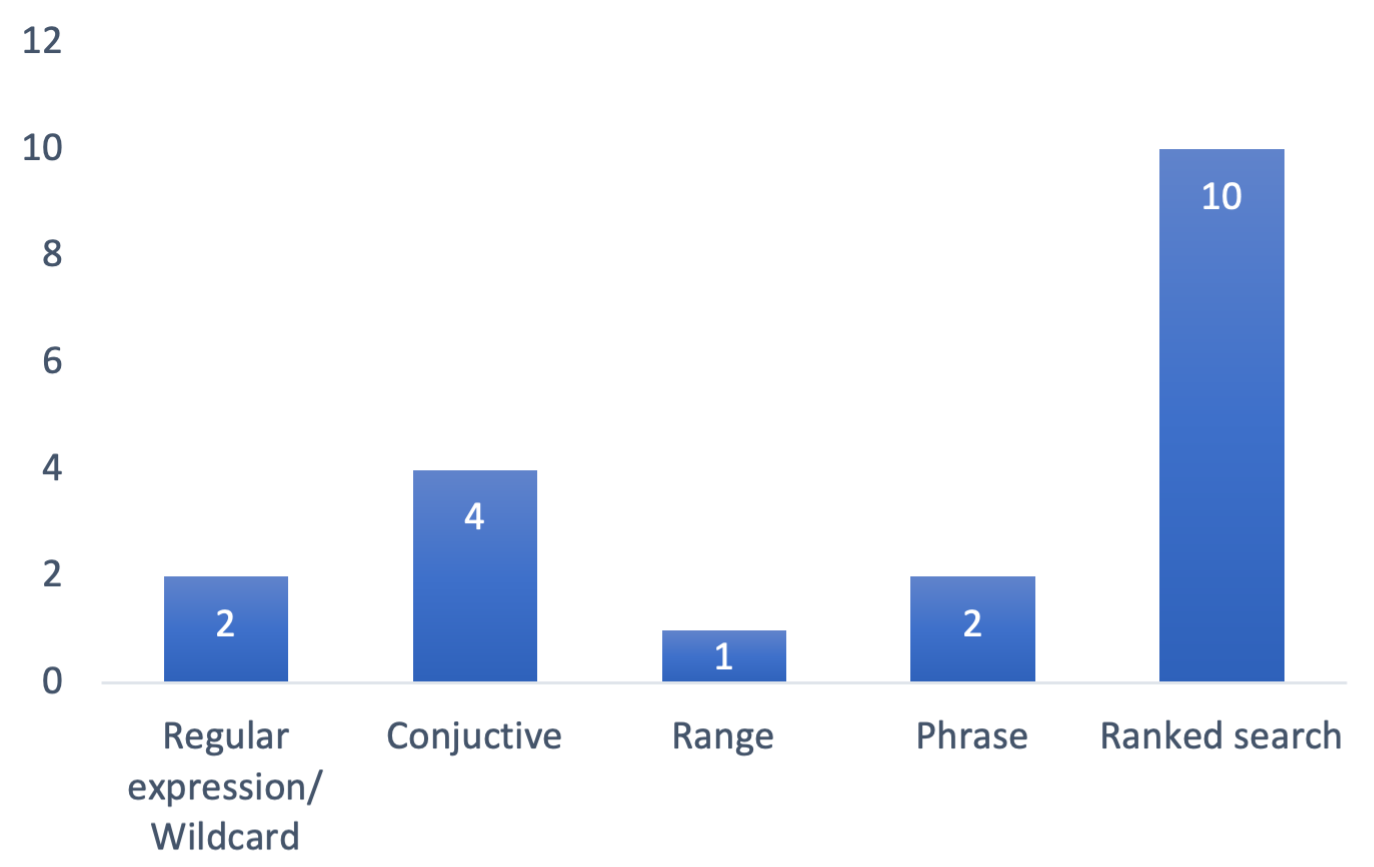}
\caption{Uses of HE on Search functionalities}
\label{Figure5}
\end{figure}

The usage of HE to improve or allow for search functionalities is very prevalent in the selected works. There is, however, a functionality which is not present in none of the works, namely the ability to perform Fuzzy Keyword search.  Figure~\ref{Figure5} illustrates the number of works that support each search functionality, considering the functionalities that appeared at least once.

From our analysis, we observed that ranked search is the most prevailing search functionality, being present in 10 out of the 23 analyzed works. Moreover, in all of those works, this functionality is achieved using the homomorphic properties of the underlying HE scheme. The most commonly used protocol involves ranking documents according to TF-IDF, where the information needed to compute relevance scores is encrypted using HE. Therefore, relevance scores can be computed in the public cloud. One of these schemes, FASE~\citeyearpar{liu_fase_2022}, which allows for multi-keyword queries, firstly ranks documents by keyword matching degree, i.e., how many queried keywords are present in the documents, then the secondary criteria are the respective relevance scores. Among all studied schemes, there is only one that uses a technique other than TF-IDF to rank their search results, namely the double score weighting formula (firstly proposed by \citet{boucennaConceptbasedSemanticSearch2016} and used in the work of Boucenna et al.~\citeyearpar{boucenna_secure_2019}).\\
Another important observation, now regarding schemes that allow for top-$k$ results, is that the majority of these schemes do not allow it in a direct and single round of communication, i.e., most schemes either rely on PIR~(\citealt{tosun_fsds_2021}, \citealt{yang_multi-user_2020}), or they include another entity in the framework, such as a coprocessor~(\citealt{elizabeth_verifiable_2020}, \citealt{elizabeth_tsed_2018}), a collaborate server~(\citealt{li_efficient_2020}) or a private cloud~(\citealt{zhang_secure_2020}), which have access to the private key and are responsible for ranking search results.\\

Regarding the multiplicity of keywords, we observed that the great majority supports multi-keyword queries. There are, however, some articles that only support single keyword queries~(\citealt{malik_homomorphic_2023}, \citealt{iqbal_novel_2022}, \citealt{gan_towards_2022}, \citealt{zhang_secure_2020}, \citealt{akavia_secure_2018}). Moreover, the ones that support multi-keyword queries can be further split into two distinct categories: multi-keyword aimed at ranked search and multi-keyword meant for conjunctive or disjunctive queries. 

Some of the articles allow only for conjunctive queries~(\citealt{wang_achieving_2022}, \citealt{wu_verifiable_2018}, \citealt{hou_privacy-preserving_2021}, \citealt{shen_secure_2019}), where the latter two also allow for phrase search, which basically is a conjunctive query where the order of keywords matters. Disjunctive queries are allowed in two of the studied works, namely the one published by Yin et al.~\citeyearpar{yin_achieve_2021} and the one by Yang et al.~\citeyearpar{yang_flexible_2020}. These approaches also allow conjunctive queries. Additionally, they allow for wildcards in each of the queried keywords, where the latter has a protocol to return the top-k results when the query performed is disjunctive. Besides the above-mentioned work~(\citealt{yang_flexible_2020}), some other schemes allow for multi-keyword queries specifically for ranked search~(\citealt{liu_fase_2022}, \citealt{andola_secure_2022}, \citealt{tosun_fsds_2021}, \citealt{elizabeth_verifiable_2020}, \citealt{li_efficient_2020}, \citealt{yang_multi-user_2020}, \citealt{boucenna_secure_2019}). From all these papers, only the VSED scheme~(\citealt{elizabeth_verifiable_2020}) has a slightly different approach on the ranking protocol. More specifically, before ranking the documents, it performs a disjunctive keyword search, to guarantee that every returned document has at least one queried keyword.
%%%%%%%%%%%%%%%%%%%%%%%%%
\subsection{HE usage in Other Functionalities}
HE has also been explored to provide functionalities such as dynamic updates and verifiability. However, during our study,  we did not find any work that mentioned the ability to delegate the search. Moreover, only two works allow the revocation of users, and they do not use HE for this purpose.

Among the properties within this category, ``Dynamic'' is the most frequent, and is always achieved using the properties of HE. Implementing dynamic updates is challenging, especially on index-based schemes, since they required some sort of index updates. Nonetheless, schemes that utilize HE to encrypt the index offer the advantage of performing computations directly on ciphertexts, thereby facilitating efficient index updates.

From the studied articles that allow for dynamic updates, half of them rely on an inverted index~(\citealt{elizabeth_verifiable_2020}, \citealt{elizabeth_tsed_2018}, \citealt{li_efficient_2020}). Hou et al.~\citeyearpar{hou_privacy-preserving_2021} uses a VBTree, which works as a tree index where the search is performed over keywords, similar to an inverted index. The work of Prakash et al.~\citeyearpar{prakash_pindex_2021} relies on a novel index type called the multi-linked index.

Regarding the ability to update, most of the approaches allow only for updates of keywords over files. However, Prakash et al.~\citeyearpar{prakash_pindex_2021} and Li et al.~\citeyearpar{li_efficient_2020} also allow for dynamic updates of files, besides allowing for regular updates over keywords.

The capability to authorize or revoke users is crucial for ensuring data privacy, yet it is not commonly addressed in the selected articles. Only two of them incorporate this property~(\citealt{yang_flexible_2020}, \citeyear{ yang_multi-user_2020}), and neither of them exploits HE schemes to achieve this functionality. Both articles follow a similar protocol, involving the generation of a search token with an expiration date. Additionally, they allow users to query multiple DOs simultaneously by requesting an authorization token from all DOs at the same time.

%%%%%%%%%%%%%%%%%%%%%%%%%
\subsection{Multiplicity of Users}
\begin{figure}[ht]
\centering
\includegraphics[width=0.8\columnwidth]{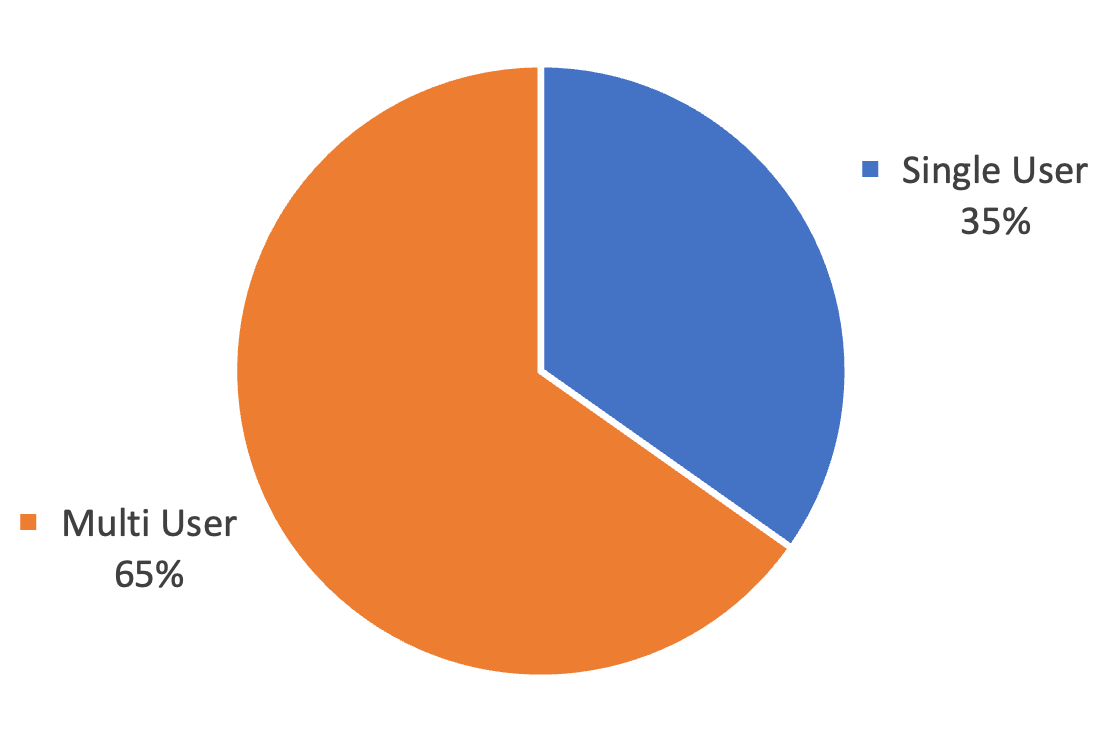}
\caption{Multiplicity of users}
\label{Figure6}
\end{figure}

The multiplicity of users is another property that we have analyzed in the selected works. Even though this property is not directly achieved using HE, we observed that schemes meant for the multi-user setting are the most common, with a total of 63\% allowing this functionality (Figure \ref{Figure6}). Nonetheless, there are a couple of papers that show different scenarios where single-user SE schemes are suitable. Malik et al. \citeyearpar{malik_homomorphic_2023}, presented an application where the single entity is an airport, and Iqbal et al.~\citeyearpar{iqbal_novel_2022} presented a use case where both the DO and DU are the same hospital.

Regarding the schemes that are meant for a multi-user setting, the majority rely on user authorization~(\citealt{liu_fase_2022}, \citealt{andola_secure_2022}, \citealt{tosun_fsds_2021}, \citealt{yang_flexible_2020}, \citealt{yang_multi-user_2020}, \citealt{boucenna_secure_2019}, \citealt{shen_secure_2019}, \citealt{wu_verifiable_2018}). Note also that Yang et al.~\citeyearpar{yang_multi-user_2020} proposed a multi-user SE scheme that allows DUs to perform search queries over the data of multiple DOs simultaneously.
From the remaining schemes, it is worth highlighting the work of Gan et al.~\citeyearpar{gan_towards_2022}, which achieves the multi-user functionality by allowing users to search through a private link, a public search tree, or both.
%%%%%%%%%%%%%%%%%%%%%%%%%%%%%%%%%%%%%%%%%%5

\section{Conclusion and Future work}\label{sec:conclusion}

In this work, we  provided a comprehensive analysis of the research trends on SE schemes that use HE. We focused our study on the types of HE schemes used, their application to enhance search structures, to allow several functionalities such as ranked, conjunctive, disjunctive, range, and phrase search, as well as verifiability, dynamic updates, and  the ability to add and revoke users. This analysis was conducted on a carefully selected set of 23 works, following a well-defined research methodology

Our findings revealed that HE usage in SE schemes has increased in recent years, with 75\% of the selected works published within the last three years. The most commonly used type of HE schemes in SE is PHE, which accounted for the majority of the analysed schemes. The popularity of PHE schemes, particularly the widespread adoption of Paillier's cryptosystem, can be attributed to its simplicity, proven security properties, and availability in several open-source libraries, making it easily accessible for researchers.

When considering the usage of HE in search structures, we found that building the index using HE is the most prevalent approach. This is not surprising, since SE schemes that rely on sequential scans are not suitable for large databases. Additionally, we analysed the types of indexes used in these schemes and observed that normal indexes are the most common choice, followed by inverted indexes and tree indexes.

Regarding the usage of HE in search functionalities, we found that ranked search is the most prevalent functionality. Relevance scores are computed using homomorphic properties, and are often based on TF-IDF ranking. Multi-keyword queries are widely supported, either for ranked search or conjunctive/disjunctive queries. However, fuzzy keyword search functionality was not found in any of the analysed works.

In future research, there is a need to explore and improve upon various functionalities when searching over encrypted data using HE schemes. While dynamic updates consistently take advantage of homomorphic properties, other functionalities such as authorizing and revoking users, verifiability, and delegation of search capability require further development. Among the analysed works, only one study~(\citealt{wu_verifiable_2018}) utilized a HE scheme to achieve verifiability, while none of the articles addressed the ability to authorize and revoke users.  However, we found two works that, although not employing HE schemes directly, utilized a cryptographic primitive called Homomorphic Message Authentication to facilitate verifiability (Zhang et al., 2020; Wan et al., 2018).  Furthermore, Zhang et al.'s work utilized this primitive to enable the authorization and revocation of users' access. Therefore, considering the potential benefits and the limited exploration in this area, it is worth further exploring and investigating the utilization of Homomorphic Message Authentication and similar cryptographic primitives for achieving verifiability, and the ability to add and revoke users.

Although FHE schemes were present in nearly half of the studied articles, we believe that the percentage could be significantly higher given the recent advancements in FHE schemes. For example, the TFHE scheme, introduced in \citeyear{chillotti_tfhe_2020}, is an FHE scheme worth exploring.  Leveraging other encryption schemes alongside HE, such as Attribute-Based Encryption used for user access control (e.g. it is used in the work of Boucenna et al.,~\citeyear{boucenna_secure_2019}), or OPE for facilitating ciphertext comparison (e.g. it is used in FASE, \citeyear{liu_fase_2022}), could lead to the development of more sophisticated schemes.

In our study, we also concluded that the studied articles cover several important search functionalities. However, there were some notable omissions, particularly in the areas of fuzzy keyword search and occurrence queries. Fuzzy keyword search is one of the most desirable functionalities in real-life applications, making it crucial to explore HE schemes that can support this functionality. One approach, as suggested by Dong et al.~\citeyearpar{dong_fuzzy_2013}, involves using Hamming distances. Additionally, occurrence queries were not addressed in any of the studied schemes, but we observed that most articles allowing ranked searches could easily accommodate occurrence queries since they already require information such as TF to rank documents.

Finally, it is important for future research to focus on reducing the reliance on retrieval protocols like PIR. The emphasis should be on developing SE schemes that operate effectively with a minimum number of communication rounds, aiming to minimize the overall overhead.

\section*{Declaration of competing interest}
The authors declare that they have no competing interests.

\section*{Data availability}
No data was used for the research described in the article.

\section*{Acknowledgements}%% if any
This work was partially supported by the Norte Portugal Regional Operational Programme (NORTE 2020), under the PORTUGAL 2020 Partnership Agreement, through the European Regional Development Fund (ERDF), within project ``Cybers SeC IP'' (NORTE-01-0145-FEDER-000044).

\bibliographystyle{cas-model2-names}

\bibliography{cas-refs}

\begin{thebibliography}{81}
\expandafter\ifx\csname natexlab\endcsname\relax\def\natexlab#1{#1}\fi
\providecommand{\url}[1]{\texttt{#1}}
\providecommand{\href}[2]{#2}
\providecommand{\path}[1]{#1}
\providecommand{\DOIprefix}{doi:}
\providecommand{\ArXivprefix}{arXiv:}
\providecommand{\URLprefix}{URL: }
\providecommand{\Pubmedprefix}{pmid:}
\providecommand{\doi}[1]{\href{http://dx.doi.org/#1}{\path{#1}}}
\providecommand{\Pubmed}[1]{\href{pmid:#1}{\path{#1}}}
\providecommand{\bibinfo}[2]{#2}
\ifx\xfnm\relax \def\xfnm[#1]{\unskip,\space#1}\fi
%Type = Article
\bibitem[{Acar et~al.(2018)Acar, Aksu, Uluagac and Conti}]{acar_survey_2018}
\bibinfo{author}{Acar, A.}, \bibinfo{author}{Aksu, H.},
  \bibinfo{author}{Uluagac, A.}, \bibinfo{author}{Conti, M.},
  \bibinfo{year}{2018}.
\newblock \bibinfo{title}{A survey on homomorphic encryption schemes: {Theory}
  and implementation}.
\newblock \bibinfo{journal}{ACM Computing Surveys} \bibinfo{volume}{51}.
%Type = Article
\bibitem[{Agrawal(2021)}]{Agrawal2021}
\bibinfo{author}{Agrawal, S.}, \bibinfo{year}{2021}.
\newblock \bibinfo{title}{A survey on recent applications of cloud computing in
  education: Covid-19 perspective}.
\newblock \bibinfo{journal}{Journal of Physics: Conference Series}
  \bibinfo{volume}{1828}, \bibinfo{pages}{012076}.
%Type = Inproceedings
\bibitem[{Akavia et~al.(2018)Akavia, Feldman and Shaul}]{akavia_secure_2018}
\bibinfo{author}{Akavia, A.}, \bibinfo{author}{Feldman, D.},
  \bibinfo{author}{Shaul, H.}, \bibinfo{year}{2018}.
\newblock \bibinfo{title}{Secure {Search} on {Encrypted} {Data} via
  {Multi}-{Ring} {Sketch}}, in: \bibinfo{booktitle}{Proceedings of the 2018
  {ACM} {SIGSAC} {Conference} on {Computer} and {Communications} {Security}},
  \bibinfo{publisher}{Association for Computing Machinery},
  \bibinfo{address}{New York, NY, USA}. pp. \bibinfo{pages}{985--1001}.
%Type = Article
\bibitem[{Alloghani et~al.(2019)Alloghani, Alani, Al-Jumeily, Baker, Mustafina,
  Hussain and Aljaaf}]{alloghani_systematic_2019}
\bibinfo{author}{Alloghani, M.}, \bibinfo{author}{Alani, M.M.},
  \bibinfo{author}{Al-Jumeily, D.}, \bibinfo{author}{Baker, T.},
  \bibinfo{author}{Mustafina, J.}, \bibinfo{author}{Hussain, A.},
  \bibinfo{author}{Aljaaf, A.J.}, \bibinfo{year}{2019}.
\newblock \bibinfo{title}{A systematic review on the status and progress of
  homomorphic encryption technologies}.
\newblock \bibinfo{journal}{Journal of Information Security and Applications}
  \bibinfo{volume}{48}, \bibinfo{pages}{102362}.
%Type = Article
\bibitem[{Andola et~al.(2022a)Andola, Gahlot, Yadav, Venkatesan and
  Verma}]{andola_searchable_2022}
\bibinfo{author}{Andola, N.}, \bibinfo{author}{Gahlot, R.},
  \bibinfo{author}{Yadav, V.}, \bibinfo{author}{Venkatesan, S.},
  \bibinfo{author}{Verma, S.}, \bibinfo{year}{2022}a.
\newblock \bibinfo{title}{Searchable encryption on the cloud: a survey}.
\newblock \bibinfo{journal}{Journal of Supercomputing} \bibinfo{volume}{78},
  \bibinfo{pages}{9952--9984}.
%Type = Article
\bibitem[{Andola et~al.(2022b)Andola, Prakash, Yadav, Raghav, Venkatesan and
  Verma}]{andola_secure_2022}
\bibinfo{author}{Andola, N.}, \bibinfo{author}{Prakash, S.},
  \bibinfo{author}{Yadav, V.K.}, \bibinfo{author}{Raghav},
  \bibinfo{author}{Venkatesan, S.}, \bibinfo{author}{Verma, S.},
  \bibinfo{year}{2022}b.
\newblock \bibinfo{title}{A secure searchable encryption scheme for cloud using
  hash-based indexing}.
\newblock \bibinfo{journal}{Journal of Computer and System Sciences}
  \bibinfo{volume}{126}, \bibinfo{pages}{119--137}.
%Type = Article
\bibitem[{Bader and Michala(2021)}]{bader_searchable_2021}
\bibinfo{author}{Bader, J.}, \bibinfo{author}{Michala, A.},
  \bibinfo{year}{2021}.
\newblock \bibinfo{title}{Searchable {Encryption} with {Access} {Control} in
  {Industrial} {Internet} of {Things} ({IIoT})}.
\newblock \bibinfo{journal}{Wireless Communications and Mobile Computing}
  \bibinfo{volume}{2021}.
%Type = Inproceedings
\bibitem[{Benaloh(1994)}]{benaloh_dense_1994}
\bibinfo{author}{Benaloh, J.}, \bibinfo{year}{1994}.
\newblock \bibinfo{title}{Dense probabilistic encryption}, in:
  \bibinfo{booktitle}{Proceedings of the workshop on selected areas of
  cryptography}, pp. \bibinfo{pages}{120--128}.
%Type = Inproceedings
\bibitem[{Boneh et~al.(2005)Boneh, Goh and Nissim}]{boneh_evaluating_2005}
\bibinfo{author}{Boneh, D.}, \bibinfo{author}{Goh, E.J.},
  \bibinfo{author}{Nissim, K.}, \bibinfo{year}{2005}.
\newblock \bibinfo{title}{Evaluating 2-{DNF} {Formulas} on {Ciphertexts}}, in:
  \bibinfo{editor}{Kilian, J.} (Ed.), \bibinfo{booktitle}{Theory of
  {Cryptography}}, \bibinfo{publisher}{Springer}, \bibinfo{address}{Berlin,
  Heidelberg}. pp. \bibinfo{pages}{325--341}.
%Type = Inproceedings
\bibitem[{Boneh et~al.(2007)Boneh, Kushilevitz, Ostrovsky and
  Skeith}]{boneh_public_2007}
\bibinfo{author}{Boneh, D.}, \bibinfo{author}{Kushilevitz, E.},
  \bibinfo{author}{Ostrovsky, R.}, \bibinfo{author}{Skeith, W.E.},
  \bibinfo{year}{2007}.
\newblock \bibinfo{title}{Public {Key} {Encryption} {That} {Allows} {PIR}
  {Queries}}, in: \bibinfo{editor}{Menezes, A.} (Ed.),
  \bibinfo{booktitle}{Advances in {Cryptology} - {CRYPTO} 2007},
  \bibinfo{publisher}{Springer}, \bibinfo{address}{Berlin, Heidelberg}. pp.
  \bibinfo{pages}{50--67}.
%Type = Inproceedings
\bibitem[{Boneh and Waters(2007)}]{boneh_conjunctive_2007}
\bibinfo{author}{Boneh, D.}, \bibinfo{author}{Waters, B.},
  \bibinfo{year}{2007}.
\newblock \bibinfo{title}{Conjunctive, {Subset}, and {Range} {Queries} on
  {Encrypted} {Data}}, in: \bibinfo{editor}{Vadhan, S.P.} (Ed.),
  \bibinfo{booktitle}{Theory of {Cryptography}}, \bibinfo{publisher}{Springer},
  \bibinfo{address}{Berlin, Heidelberg}. pp. \bibinfo{pages}{535--554}.
%Type = Inproceedings
\bibitem[{Boucenna et~al.(2016)Boucenna, Nouali and
  Kechid}]{boucennaConceptbasedSemanticSearch2016}
\bibinfo{author}{Boucenna, F.}, \bibinfo{author}{Nouali, O.},
  \bibinfo{author}{Kechid, S.}, \bibinfo{year}{2016}.
\newblock \bibinfo{title}{Concept-based {Semantic} {Search} over {Encrypted}
  {Cloud} {Data}:}, in: \bibinfo{booktitle}{Proceedings of the 12th
  {International} {Conference} on {Web} {Information} {Systems} and
  {Technologies}}, \bibinfo{publisher}{SCITEPRESS - Science and and Technology
  Publications}, \bibinfo{address}{Rome, Italy}. pp. \bibinfo{pages}{235--242}.
%Type = Article
\bibitem[{Boucenna et~al.(2019)Boucenna, Nouali, Kechid and
  Tahar~Kechadi}]{boucenna_secure_2019}
\bibinfo{author}{Boucenna, F.}, \bibinfo{author}{Nouali, O.},
  \bibinfo{author}{Kechid, S.}, \bibinfo{author}{Tahar~Kechadi, M.},
  \bibinfo{year}{2019}.
\newblock \bibinfo{title}{Secure {Inverted} {Index} {Based} {Search} over
  {Encrypted} {Cloud} {Data} with {User} {Access} {Rights} {Management}}.
\newblock \bibinfo{journal}{Journal of Computer Science and Technology}
  \bibinfo{volume}{34}, \bibinfo{pages}{133--154}.
%Type = Inproceedings
\bibitem[{Brakerski et~al.(2012)Brakerski, Gentry and
  Vaikuntanathan}]{brakerski_leveled_2012}
\bibinfo{author}{Brakerski, Z.}, \bibinfo{author}{Gentry, C.},
  \bibinfo{author}{Vaikuntanathan, V.}, \bibinfo{year}{2012}.
\newblock \bibinfo{title}{({Leveled}) fully homomorphic encryption without
  bootstrapping}, in: \bibinfo{booktitle}{Proceedings of the 3rd {Innovations}
  in {Theoretical} {Computer} {Science} {Conference}},
  \bibinfo{publisher}{Association for Computing Machinery},
  \bibinfo{address}{New York, NY, USA}. pp. \bibinfo{pages}{309--325}.
%Type = Inproceedings
\bibitem[{Brakerski and Vaikuntanathan(2011)}]{brakerski_fully_2011}
\bibinfo{author}{Brakerski, Z.}, \bibinfo{author}{Vaikuntanathan, V.},
  \bibinfo{year}{2011}.
\newblock \bibinfo{title}{Fully {Homomorphic} {Encryption} from {Ring}-{LWE}
  and {Security} for {Key} {Dependent} {Messages}}, in:
  \bibinfo{editor}{Rogaway, P.} (Ed.), \bibinfo{booktitle}{Advances in
  {Cryptology} – {CRYPTO} 2011}, \bibinfo{publisher}{Springer},
  \bibinfo{address}{Berlin, Heidelberg}. pp. \bibinfo{pages}{505--524}.
%Type = Inproceedings
\bibitem[{Bresson et~al.(2003)Bresson, Catalano and
  Pointcheval}]{bressonSimplePublicKeyCryptosystem2003}
\bibinfo{author}{Bresson, E.}, \bibinfo{author}{Catalano, D.},
  \bibinfo{author}{Pointcheval, D.}, \bibinfo{year}{2003}.
\newblock \bibinfo{title}{A {Simple} {Public}-{Key} {Cryptosystem} with a
  {Double} {Trapdoor} {Decryption} {Mechanism} and {Its} {Applications}}, in:
  \bibinfo{editor}{Laih, C.S.} (Ed.), \bibinfo{booktitle}{Advances in
  {Cryptology} - {ASIACRYPT} 2003}, \bibinfo{publisher}{Springer},
  \bibinfo{address}{Berlin, Heidelberg}. pp. \bibinfo{pages}{37--54}.
%Type = Article
\bibitem[{Bösch et~al.(2014)Bösch, Hartel, Jonker and
  Peter}]{bosch_survey_2014}
\bibinfo{author}{Bösch, C.}, \bibinfo{author}{Hartel, P.},
  \bibinfo{author}{Jonker, W.}, \bibinfo{author}{Peter, A.},
  \bibinfo{year}{2014}.
\newblock \bibinfo{title}{A {Survey} of {Provably} {Secure} {Searchable}
  {Encryption}}.
\newblock \bibinfo{journal}{ACM Computing Surveys} \bibinfo{volume}{47},
  \bibinfo{pages}{18:1--18:51}.
%Type = Article
\bibitem[{Cao et~al.(2014)Cao, Wang, Li, Ren and
  Lou}]{caoPrivacyPreservingMultiKeywordRanked2014}
\bibinfo{author}{Cao, N.}, \bibinfo{author}{Wang, C.}, \bibinfo{author}{Li,
  M.}, \bibinfo{author}{Ren, K.}, \bibinfo{author}{Lou, W.},
  \bibinfo{year}{2014}.
\newblock \bibinfo{title}{Privacy-{Preserving} {Multi}-{Keyword} {Ranked}
  {Search} over {Encrypted} {Cloud} {Data}}.
\newblock \bibinfo{journal}{IEEE Transactions on Parallel and Distributed
  Systems} \bibinfo{volume}{25}, \bibinfo{pages}{222--233}.
%Type = Article
\bibitem[{Challa(2020)}]{challa_homomorphic_2020}
\bibinfo{author}{Challa, R.}, \bibinfo{year}{2020}.
\newblock \bibinfo{title}{Homomorphic {Encryption}: {Review} and
  {Applications}}.
\newblock \bibinfo{journal}{Lecture Notes on Data Engineering and
  Communications Technologies} \bibinfo{volume}{37}, \bibinfo{pages}{273--281}.
%Type = Inproceedings
\bibitem[{Cheon et~al.(2017)Cheon, Kim, Kim and Song}]{cheon_homomorphic_2017}
\bibinfo{author}{Cheon, J.H.}, \bibinfo{author}{Kim, A.}, \bibinfo{author}{Kim,
  M.}, \bibinfo{author}{Song, Y.}, \bibinfo{year}{2017}.
\newblock \bibinfo{title}{Homomorphic {Encryption} for {Arithmetic} of
  {Approximate} {Numbers}}, in: \bibinfo{editor}{Takagi, T.},
  \bibinfo{editor}{Peyrin, T.} (Eds.), \bibinfo{booktitle}{Advances in
  {Cryptology} – {ASIACRYPT} 2017}, \bibinfo{publisher}{Springer
  International Publishing}, \bibinfo{address}{Cham}. pp.
  \bibinfo{pages}{409--437}.
%Type = Article
\bibitem[{Chillotti et~al.(2020)Chillotti, Gama, Georgieva and
  Izabachène}]{chillotti_tfhe_2020}
\bibinfo{author}{Chillotti, I.}, \bibinfo{author}{Gama, N.},
  \bibinfo{author}{Georgieva, M.}, \bibinfo{author}{Izabachène, M.},
  \bibinfo{year}{2020}.
\newblock \bibinfo{title}{{TFHE}: {Fast} {Fully} {Homomorphic} {Encryption}
  {Over} the {Torus}}.
\newblock \bibinfo{journal}{Journal of Cryptology} \bibinfo{volume}{33},
  \bibinfo{pages}{34--91}.
%Type = Inproceedings
\bibitem[{Choi et~al.(2021)Choi, Dachman-Soled, Gordon, Liu and
  Yerukhimovich}]{choi_compressed_2021}
\bibinfo{author}{Choi, S.G.}, \bibinfo{author}{Dachman-Soled, D.},
  \bibinfo{author}{Gordon, S.D.}, \bibinfo{author}{Liu, L.},
  \bibinfo{author}{Yerukhimovich, A.}, \bibinfo{year}{2021}.
\newblock \bibinfo{title}{Compressed {Oblivious} {Encoding} for
  {Homomorphically} {Encrypted} {Search}}, in: \bibinfo{booktitle}{Proceedings
  of the 2021 {ACM} {SIGSAC} {Conference} on {Computer} and {Communications}
  {Security}}, \bibinfo{publisher}{Association for Computing Machinery},
  \bibinfo{address}{New York, NY, USA}. pp. \bibinfo{pages}{2277--2291}.
%Type = Inproceedings
\bibitem[{van Dijk et~al.(2010)van Dijk, Gentry, Halevi and
  Vaikuntanathan}]{van_dijk_fully_2010}
\bibinfo{author}{van Dijk, M.}, \bibinfo{author}{Gentry, C.},
  \bibinfo{author}{Halevi, S.}, \bibinfo{author}{Vaikuntanathan, V.},
  \bibinfo{year}{2010}.
\newblock \bibinfo{title}{Fully {Homomorphic} {Encryption} over the
  {Integers}}, in: \bibinfo{editor}{Gilbert, H.} (Ed.),
  \bibinfo{booktitle}{Advances in {Cryptology} – {EUROCRYPT} 2010},
  \bibinfo{publisher}{Springer}, \bibinfo{address}{Berlin, Heidelberg}. pp.
  \bibinfo{pages}{24--43}.
%Type = Article
\bibitem[{Dong et~al.(2013)Dong, Guan, Wu and Chen}]{dong_fuzzy_2013}
\bibinfo{author}{Dong, Q.}, \bibinfo{author}{Guan, Z.}, \bibinfo{author}{Wu,
  L.}, \bibinfo{author}{Chen, Z.}, \bibinfo{year}{2013}.
\newblock \bibinfo{title}{Fuzzy keyword search over encrypted data in the
  public key setting}.
\newblock \bibinfo{journal}{Lecture Notes in Computer Science (including
  subseries Lecture Notes in Artificial Intelligence and Lecture Notes in
  Bioinformatics)} \bibinfo{volume}{7923 LNCS}, \bibinfo{pages}{729--740}.
%Type = Article
\bibitem[{Dowsley et~al.(2017)Dowsley, Michalas, Nagel and
  Paladi}]{dowsley_survey_2017}
\bibinfo{author}{Dowsley, R.}, \bibinfo{author}{Michalas, A.},
  \bibinfo{author}{Nagel, M.}, \bibinfo{author}{Paladi, N.},
  \bibinfo{year}{2017}.
\newblock \bibinfo{title}{A survey on design and implementation of protected
  searchable data in the cloud}.
\newblock \bibinfo{journal}{Computer Science Review} \bibinfo{volume}{26},
  \bibinfo{pages}{17--30}.
%Type = Misc
\bibitem[{Dworkin et~al.(2001)Dworkin, Barker, Nechvatal, Foti, Bassham, Roback
  and Dray}]{117146}
\bibinfo{author}{Dworkin, M.}, \bibinfo{author}{Barker, E.},
  \bibinfo{author}{Nechvatal, J.}, \bibinfo{author}{Foti, J.},
  \bibinfo{author}{Bassham, L.}, \bibinfo{author}{Roback, E.},
  \bibinfo{author}{Dray, J.}, \bibinfo{year}{2001}.
\newblock \bibinfo{title}{Advanced {E}ncryption {S}tandard ({AES})}.
\newblock \URLprefix \url{https://doi.org/10.6028/NIST.FIPS.197}.
  \bibinfo{note}{{A}ccessed May 22 2023}.
%Type = Article
\bibitem[{Elgamal(1985)}]{elgamal_public_1985}
\bibinfo{author}{Elgamal, T.}, \bibinfo{year}{1985}.
\newblock \bibinfo{title}{A public key cryptosystem and a signature scheme
  based on discrete logarithms}.
\newblock \bibinfo{journal}{IEEE Transactions on Information Theory}
  \bibinfo{volume}{31}, \bibinfo{pages}{469--472}.
%Type = Article
\bibitem[{Elizabeth and Prakash(2020)}]{elizabeth_verifiable_2020}
\bibinfo{author}{Elizabeth, B.}, \bibinfo{author}{Prakash, A.},
  \bibinfo{year}{2020}.
\newblock \bibinfo{title}{Verifiable top-k searchable encryption for cloud
  data}.
\newblock \bibinfo{journal}{Sadhana - Academy Proceedings in Engineering
  Sciences} \bibinfo{volume}{45}.
%Type = Article
\bibitem[{Elizabeth et~al.(2018)Elizabeth, Prakash and
  Uthariaraj}]{elizabeth_tsed_2018}
\bibinfo{author}{Elizabeth, B.}, \bibinfo{author}{Prakash, A.},
  \bibinfo{author}{Uthariaraj, V.}, \bibinfo{year}{2018}.
\newblock \bibinfo{title}{{TSED}: {Top}-k ranked searchable encryption for
  secure cloud data storage}.
\newblock \bibinfo{journal}{Advances in Intelligent Systems and Computing}
  \bibinfo{volume}{645}, \bibinfo{pages}{113--121}.
%Type = Misc
\bibitem[{Fan and Vercauteren(2012)}]{fan_somewhat_2012}
\bibinfo{author}{Fan, J.}, \bibinfo{author}{Vercauteren, F.},
  \bibinfo{year}{2012}.
\newblock \bibinfo{title}{Somewhat practical fully homomorphic encryption}.
\newblock \bibinfo{howpublished}{Cryptology ePrint Archive, Paper 2012/144}.
\newblock \URLprefix \url{https://eprint.iacr.org/2012/144}.
  \bibinfo{note}{{A}ccessed Feb 24 2023}.
%Type = Article
\bibitem[{Gan et~al.(2022)Gan, Wang, Huang, Li, Zhou and
  Wang}]{gan_towards_2022}
\bibinfo{author}{Gan, Q.}, \bibinfo{author}{Wang, X.}, \bibinfo{author}{Huang,
  D.}, \bibinfo{author}{Li, J.}, \bibinfo{author}{Zhou, D.},
  \bibinfo{author}{Wang, C.}, \bibinfo{year}{2022}.
\newblock \bibinfo{title}{Towards multi-client forward private searchable
  symmetric encryption in cloud computing}.
\newblock \bibinfo{journal}{IEEE Transactions on Services Computing}
  \bibinfo{volume}{15}, \bibinfo{pages}{3566--3576}.
%Type = Phdthesis
\bibitem[{Gentry(2009)}]{gentry_fully_2009}
\bibinfo{author}{Gentry, C.}, \bibinfo{year}{2009}.
\newblock \bibinfo{title}{A {Fully} {Homomorphic} {Encryption} {Scheme}}.
\newblock \bibinfo{type}{{PhD} {Thesis}}. Stanford University.
  \bibinfo{address}{Stanford, CA, USA}.
%Type = Inproceedings
\bibitem[{Gentry and Halevi(2011)}]{gentry_implementing_2011}
\bibinfo{author}{Gentry, C.}, \bibinfo{author}{Halevi, S.},
  \bibinfo{year}{2011}.
\newblock \bibinfo{title}{Implementing {Gentry}’s {Fully}-{Homomorphic}
  {Encryption} {Scheme}}, in: \bibinfo{editor}{Paterson, K.G.} (Ed.),
  \bibinfo{booktitle}{Advances in {Cryptology} – {EUROCRYPT} 2011},
  \bibinfo{publisher}{Springer}, \bibinfo{address}{Berlin, Heidelberg}. pp.
  \bibinfo{pages}{129--148}.
\newblock \DOIprefix\doi{10.1007/978-3-642-20465-4_9}.
%Type = Inproceedings
\bibitem[{Goldwasser and Micali(1982)}]{goldwasser_probabilistic_1982}
\bibinfo{author}{Goldwasser, S.}, \bibinfo{author}{Micali, S.},
  \bibinfo{year}{1982}.
\newblock \bibinfo{title}{Probabilistic encryption \&amp; how to play mental
  poker keeping secret all partial information}, in:
  \bibinfo{booktitle}{Proceedings of the fourteenth annual {ACM} symposium on
  {Theory} of computing}, \bibinfo{publisher}{Association for Computing
  Machinery}, \bibinfo{address}{New York, NY, USA}. pp.
  \bibinfo{pages}{365--377}.
%Type = Article
\bibitem[{Gonz\'alez-Mart\'inez et~al.(2015)Gonz\'alez-Mart\'inez,
  Bote-Lorenzo, G\'omez-S\'anchez and Cano-Parra}]{Gonzalez-Martinez2015132}
\bibinfo{author}{Gonz\'alez-Mart\'inez, J.A.}, \bibinfo{author}{Bote-Lorenzo,
  M.L.}, \bibinfo{author}{G\'omez-S\'anchez, E.}, \bibinfo{author}{Cano-Parra,
  R.}, \bibinfo{year}{2015}.
\newblock \bibinfo{title}{Cloud computing and education: A state-of-the-art
  survey}.
\newblock \bibinfo{journal}{Computers and Education} \bibinfo{volume}{80},
  \bibinfo{pages}{132 – 151}.
%Type = Article
\bibitem[{Guo et~al.(2019)Guo, Zhuang, Jie, Choo and Tang}]{guo_secure_2019}
\bibinfo{author}{Guo, C.}, \bibinfo{author}{Zhuang, R.}, \bibinfo{author}{Jie,
  Y.}, \bibinfo{author}{Choo, K.K.}, \bibinfo{author}{Tang, X.},
  \bibinfo{year}{2019}.
\newblock \bibinfo{title}{Secure range search over encrypted uncertain {IoT}
  outsourced data}.
\newblock \bibinfo{journal}{IEEE Internet of Things Journal}
  \bibinfo{volume}{6}, \bibinfo{pages}{1520--1529}.
%Type = Inproceedings
\bibitem[{Halevi and Shoup(2014)}]{haleviAlgorithmsHElib2014}
\bibinfo{author}{Halevi, S.}, \bibinfo{author}{Shoup, V.},
  \bibinfo{year}{2014}.
\newblock \bibinfo{title}{Algorithms in {HElib}}, in: \bibinfo{editor}{Garay,
  J.A.}, \bibinfo{editor}{Gennaro, R.} (Eds.), \bibinfo{booktitle}{Advances in
  {Cryptology} – {CRYPTO} 2014}, \bibinfo{publisher}{Springer},
  \bibinfo{address}{Berlin, Heidelberg}. pp. \bibinfo{pages}{554--571}.
%Type = Article
\bibitem[{Han et~al.(2016)Han, Qin and Hu}]{han_secure_2016}
\bibinfo{author}{Han, F.}, \bibinfo{author}{Qin, J.}, \bibinfo{author}{Hu, J.},
  \bibinfo{year}{2016}.
\newblock \bibinfo{title}{Secure searches in the cloud: {A} survey}.
\newblock \bibinfo{journal}{Future Generation Computer Systems}
  \bibinfo{volume}{62}, \bibinfo{pages}{66--75}.
%Type = Article
\bibitem[{Handa et~al.(2019)Handa, Krishna and
  Aggarwal}]{handa_searchable_2019}
\bibinfo{author}{Handa, R.}, \bibinfo{author}{Krishna, C.R.},
  \bibinfo{author}{Aggarwal, N.}, \bibinfo{year}{2019}.
\newblock \bibinfo{title}{Searchable encryption: {A} survey on
  privacy-preserving search schemes on encrypted outsourced data}.
\newblock \bibinfo{journal}{Concurrency and Computation: Practice and
  Experience} \bibinfo{volume}{31}, \bibinfo{pages}{e5201}.
%Type = Inproceedings
\bibitem[{Hou et~al.(2022)Hou, Liu and Hao}]{hou_privacy-preserving_2021}
\bibinfo{author}{Hou, J.}, \bibinfo{author}{Liu, Y.}, \bibinfo{author}{Hao,
  R.}, \bibinfo{year}{2022}.
\newblock \bibinfo{title}{Privacy-preserving phrase search over encrypted
  data}, in: \bibinfo{booktitle}{Proceedings of the 4th International
  Conference on Big Data Technologies}, \bibinfo{publisher}{Association for
  Computing Machinery}, \bibinfo{address}{New York, NY, USA}. p.
  \bibinfo{pages}{154–159}.
%Type = Article
\bibitem[{How and Heng(2022)}]{how_blockchain-enabled_2022}
\bibinfo{author}{How, H.B.}, \bibinfo{author}{Heng, S.H.},
  \bibinfo{year}{2022}.
\newblock \bibinfo{title}{Blockchain-{Enabled} {Searchable} {Encryption} in
  {Clouds}: {A} {Review}}.
\newblock \bibinfo{journal}{Journal of Information Security and Applications}
  \bibinfo{volume}{67}, \bibinfo{pages}{103183}.
%Type = Article
\bibitem[{Iqbal et~al.(2022)Iqbal, Tahir, Tahir, Khan, Saeed, Almuhaideb and
  Syed}]{iqbal_novel_2022}
\bibinfo{author}{Iqbal, Y.}, \bibinfo{author}{Tahir, S.},
  \bibinfo{author}{Tahir, H.}, \bibinfo{author}{Khan, F.},
  \bibinfo{author}{Saeed, S.}, \bibinfo{author}{Almuhaideb, A.M.},
  \bibinfo{author}{Syed, A.M.}, \bibinfo{year}{2022}.
\newblock \bibinfo{title}{A novel homomorphic approach for preserving privacy
  of patient data in telemedicine}.
\newblock \bibinfo{journal}{Sensors} \bibinfo{volume}{22}.
%Type = Misc
\bibitem[{Krawczyk et~al.(1997)Krawczyk, Bellare and Canetti}]{rfc2104}
\bibinfo{author}{Krawczyk, D.H.}, \bibinfo{author}{Bellare, M.},
  \bibinfo{author}{Canetti, R.}, \bibinfo{year}{1997}.
\newblock \bibinfo{title}{{HMAC: Keyed-Hashing for Message Authentication}}.
\newblock \bibinfo{howpublished}{RFC 2104}.
%Type = Article
\bibitem[{Li et~al.(2020)Li, Zhou, Xu and Ge}]{li_efficient_2020}
\bibinfo{author}{Li, Y.}, \bibinfo{author}{Zhou, F.}, \bibinfo{author}{Xu, Z.},
  \bibinfo{author}{Ge, Y.}, \bibinfo{year}{2020}.
\newblock \bibinfo{title}{An efficient two-server ranked dynamic searchable
  encryption scheme}.
\newblock \bibinfo{journal}{IEEE Access} \bibinfo{volume}{8},
  \bibinfo{pages}{86328--86344}.
\newblock \DOIprefix\doi{10.1109/ACCESS.2020.2992773}.
%Type = Article
\bibitem[{Liu et~al.(2022)Liu, Yang, Bai, Wang and Xiang}]{liu_fase_2022}
\bibinfo{author}{Liu, G.}, \bibinfo{author}{Yang, G.}, \bibinfo{author}{Bai,
  S.}, \bibinfo{author}{Wang, H.}, \bibinfo{author}{Xiang, Y.},
  \bibinfo{year}{2022}.
\newblock \bibinfo{title}{Fase: A fast and accurate privacy-preserving
  multi-keyword top-k retrieval scheme over encrypted cloud data}.
\newblock \bibinfo{journal}{IEEE Transactions on Services Computing}
  \bibinfo{volume}{15}, \bibinfo{pages}{1855--1867}.
%Type = Article
\bibitem[{Liu et~al.(2018)Liu, Yang, Wang, Xiang and Dai}]{liu_novel_2018}
\bibinfo{author}{Liu, G.}, \bibinfo{author}{Yang, G.}, \bibinfo{author}{Wang,
  H.}, \bibinfo{author}{Xiang, Y.}, \bibinfo{author}{Dai, H.},
  \bibinfo{year}{2018}.
\newblock \bibinfo{title}{A {Novel} {Secure} {Scheme} for {Supporting}
  {Complex} {SQL} {Queries} over {Encrypted} {Databases} in {Cloud}
  {Computing}}.
\newblock \bibinfo{journal}{Security and Communication Networks}
  \bibinfo{volume}{2018}, \bibinfo{pages}{e7383514}.
%Type = Article
\bibitem[{Liu et~al.(2023)Liu, Zhao, Qin, Zhang and
  Ma}]{liu_multi-keyword_2023}
\bibinfo{author}{Liu, J.}, \bibinfo{author}{Zhao, B.}, \bibinfo{author}{Qin,
  J.}, \bibinfo{author}{Zhang, X.}, \bibinfo{author}{Ma, J.},
  \bibinfo{year}{2023}.
\newblock \bibinfo{title}{Multi-{Keyword} {Ranked} {Searchable} {Encryption}
  with the {Wildcard} {Keyword} for {Data} {Sharing} in {Cloud} {Computing}}.
\newblock \bibinfo{journal}{The Computer Journal} \bibinfo{volume}{66},
  \bibinfo{pages}{184--196}.
%Type = Article
\bibitem[{López-Alt et~al.(2012)López-Alt, Tromer and
  Vaikuntanathan}]{lopez-alt_--fly_2012}
\bibinfo{author}{López-Alt, A.}, \bibinfo{author}{Tromer, E.},
  \bibinfo{author}{Vaikuntanathan, V.}, \bibinfo{year}{2012}.
\newblock \bibinfo{title}{On-the-fly multiparty computation on the cloud via
  multikey fully homomorphic encryption}.
\newblock \bibinfo{journal}{Proceedings of the forty-fourth annual ACM
  symposium on Theory of computing} , \bibinfo{pages}{1219--1234}.
%Type = Article
\bibitem[{Malik et~al.(2023)Malik, Tahir, Tahir, Ihtasham and
  Khan}]{malik_homomorphic_2023}
\bibinfo{author}{Malik, H.}, \bibinfo{author}{Tahir, S.},
  \bibinfo{author}{Tahir, H.}, \bibinfo{author}{Ihtasham, M.},
  \bibinfo{author}{Khan, F.}, \bibinfo{year}{2023}.
\newblock \bibinfo{title}{A homomorphic approach for security and privacy
  preservation of {Smart} {Airports}}.
\newblock \bibinfo{journal}{Future Generation Computer Systems}
  \bibinfo{volume}{141}, \bibinfo{pages}{500--513}.
%Type = Article
\bibitem[{Marcolla et~al.(2022)Marcolla, Sucasas, Manzano, Bassoli, Fitzek,
  Aaraj and Marcolla}]{marcolla_survey_2022}
\bibinfo{author}{Marcolla, C.}, \bibinfo{author}{Sucasas, V.},
  \bibinfo{author}{Manzano, M.}, \bibinfo{author}{Bassoli, R.},
  \bibinfo{author}{Fitzek, F.}, \bibinfo{author}{Aaraj, N.},
  \bibinfo{author}{Marcolla, C.}, \bibinfo{year}{2022}.
\newblock \bibinfo{title}{Survey on {Fully} {Homomorphic} {Encryption},
  {Theory}, and {Applications}}.
\newblock \bibinfo{journal}{Proceedings of the IEEE} \bibinfo{volume}{110},
  \bibinfo{pages}{1572--1609}.
%Type = Book
\bibitem[{Menezes et~al.(1996)Menezes, Vanstone and Oorschot}]{menezes_1996}
\bibinfo{author}{Menezes, A.J.}, \bibinfo{author}{Vanstone, S.A.},
  \bibinfo{author}{Oorschot, P.C.V.}, \bibinfo{year}{1996}.
\newblock \bibinfo{title}{Handbook of Applied Cryptography}.
\newblock \bibinfo{edition}{1st} ed., \bibinfo{publisher}{CRC Press, Inc.},
  \bibinfo{address}{USA}.
%Type = Incollection
\bibitem[{Moumtzoglou et~al.(2014)Moumtzoglou, Kastania, Ghosh, Papapanagiotou
  and Boloor}]{Ghosh20141}
\bibinfo{author}{Moumtzoglou, A.}, \bibinfo{author}{Kastania, A.N.},
  \bibinfo{author}{Ghosh, R.}, \bibinfo{author}{Papapanagiotou, I.},
  \bibinfo{author}{Boloor, K.}, \bibinfo{year}{2014}.
\newblock \bibinfo{title}{A survey on research initiatives for healthcare
  clouds}, in: \bibinfo{booktitle}{Cloud Computing Applications for Quality
  Health Care Delivery}. \bibinfo{publisher}{IGI Global},
  \bibinfo{address}{Hershey, PA, USA}, pp. \bibinfo{pages}{1--18}.
%Type = Inproceedings
\bibitem[{Naccache and Stern(1998)}]{naccache_new_1998}
\bibinfo{author}{Naccache, D.}, \bibinfo{author}{Stern, J.},
  \bibinfo{year}{1998}.
\newblock \bibinfo{title}{A new public key cryptosystem based on higher
  residues}, in: \bibinfo{booktitle}{Proceedings of the 5th {ACM} conference on
  {Computer} and communications security}, \bibinfo{publisher}{ACM},
  \bibinfo{address}{San Francisco California USA}. pp. \bibinfo{pages}{59--66}.
%Type = Techreport
\bibitem[{Netwrix(2022)}]{netwrix_cloud_2022}
\bibinfo{author}{Netwrix}, \bibinfo{year}{2022}.
\newblock \bibinfo{title}{Cloud Data Security Report}.
\newblock \bibinfo{type}{Technical Report}. Netwrix.
\newblock \URLprefix
  \url{https://www.netwrix.com/download/collaterals/Netwrix{\_}Cloud{\_}Data{\_}Security{\_}Report{\_}2022.pdf}.
  \bibinfo{note}{{A}ccessed May 16 2023}.
%Type = Article
\bibitem[{Noorallahzade et~al.(2022)Noorallahzade, Alimoradi and
  Gholami}]{noorallahzade_survey_2022}
\bibinfo{author}{Noorallahzade, M.}, \bibinfo{author}{Alimoradi, R.},
  \bibinfo{author}{Gholami, A.}, \bibinfo{year}{2022}.
\newblock \bibinfo{title}{A {Survey} on {Public} {Key} {Encryption} with
  {Keyword} {Search}: {Taxonomy} and {Methods}}.
\newblock \bibinfo{journal}{International Journal of Mathematics and
  Mathematical Sciences} \bibinfo{volume}{2022}.
%Type = Inproceedings
\bibitem[{Paillier(1999)}]{paillier_public-key_1999}
\bibinfo{author}{Paillier, P.}, \bibinfo{year}{1999}.
\newblock \bibinfo{title}{Public-{Key} {Cryptosystems} {Based} on {Composite}
  {Degree} {Residuosity} {Classes}}, in: \bibinfo{editor}{Stern, J.} (Ed.),
  \bibinfo{booktitle}{Advances in {Cryptology} — {EUROCRYPT} ’99},
  \bibinfo{publisher}{Springer}, \bibinfo{address}{Berlin, Heidelberg}. pp.
  \bibinfo{pages}{223--238}.
%Type = Article
\bibitem[{Pham et~al.(2019)Pham, Woodworth and Amini~Salehi}]{pham_survey_2019}
\bibinfo{author}{Pham, H.}, \bibinfo{author}{Woodworth, J.},
  \bibinfo{author}{Amini~Salehi, M.}, \bibinfo{year}{2019}.
\newblock \bibinfo{title}{Survey on secure search over encrypted data on the
  cloud}.
\newblock \bibinfo{journal}{Concurrency and Computation: Practice and
  Experience} \bibinfo{volume}{31}, \bibinfo{pages}{e5284}.
%Type = Article
\bibitem[{Pillai and Lal(2022)}]{pillai_blockchain-based_2022}
\bibinfo{author}{Pillai, B.}, \bibinfo{author}{Lal, N.}, \bibinfo{year}{2022}.
\newblock \bibinfo{title}{Blockchain-based {Asymmetric} {Searchable}
  {Encryption}: {A} {Comprehensive} {Survey}}.
\newblock \bibinfo{journal}{International Journal of Engineering Trends and
  Technology} \bibinfo{volume}{70}, \bibinfo{pages}{355--365}.
%Type = Article
\bibitem[{Poh et~al.(2017)Poh, Chin, Yau, Choo and
  Mohamad}]{poh_searchable_2017}
\bibinfo{author}{Poh, G.S.}, \bibinfo{author}{Chin, J.J.},
  \bibinfo{author}{Yau, W.C.}, \bibinfo{author}{Choo, K.K.R.},
  \bibinfo{author}{Mohamad, M.S.}, \bibinfo{year}{2017}.
\newblock \bibinfo{title}{Searchable {Symmetric} {Encryption}: {Designs} and
  {Challenges}}.
\newblock \bibinfo{journal}{ACM Computing Surveys} \bibinfo{volume}{50},
  \bibinfo{pages}{40:1--40:37}.
%Type = Article
\bibitem[{Prakash and Elizabeth(2021)}]{prakash_pindex_2021}
\bibinfo{author}{Prakash, A.J.}, \bibinfo{author}{Elizabeth, B.L.},
  \bibinfo{year}{2021}.
\newblock \bibinfo{title}{Pindex: Private multi-linked index for encrypted
  document retrieval}.
\newblock \bibinfo{journal}{PLOS ONE} \bibinfo{volume}{16},
  \bibinfo{pages}{1--22}.
%Type = Article
\bibitem[{Rivest et~al.(1978a)Rivest, Adleman and Dertouzos}]{rivest_data_1978}
\bibinfo{author}{Rivest, R.L.}, \bibinfo{author}{Adleman, L.},
  \bibinfo{author}{Dertouzos, M.L.}, \bibinfo{year}{1978}a.
\newblock \bibinfo{title}{On data banks and privacy homomorphisms}.
\newblock \bibinfo{journal}{Foundations of Secure Computation, Academia Press}
  , \bibinfo{pages}{169--179}.
%Type = Article
\bibitem[{Rivest et~al.(1978b)Rivest, Shamir and Adleman}]{rivest_method_1978}
\bibinfo{author}{Rivest, R.L.}, \bibinfo{author}{Shamir, A.},
  \bibinfo{author}{Adleman, L.}, \bibinfo{year}{1978}b.
\newblock \bibinfo{title}{A method for obtaining digital signatures and
  public-key cryptosystems}.
\newblock \bibinfo{journal}{Communications of the ACM} \bibinfo{volume}{21},
  \bibinfo{pages}{120--126}.
%Type = Inproceedings
\bibitem[{Scholl and Smart(2011)}]{scholl_improved_2011}
\bibinfo{author}{Scholl, P.}, \bibinfo{author}{Smart, N.P.},
  \bibinfo{year}{2011}.
\newblock \bibinfo{title}{Improved {Key} {Generation} for {Gentry}’s {Fully}
  {Homomorphic} {Encryption} {Scheme}}, in: \bibinfo{editor}{Chen, L.} (Ed.),
  \bibinfo{booktitle}{Cryptography and {Coding}},
  \bibinfo{publisher}{Springer}, \bibinfo{address}{Berlin, Heidelberg}. pp.
  \bibinfo{pages}{10--22}.
%Type = Article
\bibitem[{Sharma(2023)}]{sharma_searchable_2023}
\bibinfo{author}{Sharma, D.}, \bibinfo{year}{2023}.
\newblock \bibinfo{title}{Searchable encryption : {A} survey}.
\newblock \bibinfo{journal}{Information Security Journal} \bibinfo{volume}{32},
  \bibinfo{pages}{76--119}.
%Type = Article
\bibitem[{Shen et~al.(2019)Shen, Ma, Zhu, Du and Xu}]{shen_secure_2019}
\bibinfo{author}{Shen, M.}, \bibinfo{author}{Ma, B.}, \bibinfo{author}{Zhu,
  L.}, \bibinfo{author}{Du, X.}, \bibinfo{author}{Xu, K.},
  \bibinfo{year}{2019}.
\newblock \bibinfo{title}{Secure phrase search for intelligent processing of
  encrypted data in cloud-based iot}.
\newblock \bibinfo{journal}{IEEE Internet of Things Journal}
  \bibinfo{volume}{6}, \bibinfo{pages}{1998--2008}.
%Type = Inproceedings
\bibitem[{Song et~al.(2000)Song, Wagner and Perrig}]{song_practical_2000}
\bibinfo{author}{Song, D.X.}, \bibinfo{author}{Wagner, D.},
  \bibinfo{author}{Perrig, A.}, \bibinfo{year}{2000}.
\newblock \bibinfo{title}{Practical techniques for searches on encrypted data},
  in: \bibinfo{booktitle}{Proceeding 2000 {IEEE} {Symposium} on {Security} and
  {Privacy}. {S}\&{P} 2000}, pp. \bibinfo{pages}{44--55}.
%Type = Inproceedings
\bibitem[{Suguna et~al.(2018)Suguna, Ramalakshmi, Cynthia and
  Prakash}]{8777606}
\bibinfo{author}{Suguna, M.}, \bibinfo{author}{Ramalakshmi, M.},
  \bibinfo{author}{Cynthia, J.}, \bibinfo{author}{Prakash, D.},
  \bibinfo{year}{2018}.
\newblock \bibinfo{title}{A survey on cloud and internet of things based
  healthcare diagnosis}, in: \bibinfo{booktitle}{2018 4th International
  Conference on Computing Communication and Automation (ICCCA)}, pp.
  \bibinfo{pages}{1--4}.
%Type = Article
\bibitem[{Tosun and Savaş(2021)}]{tosun_fsds_2021}
\bibinfo{author}{Tosun, T.}, \bibinfo{author}{Savaş, E.},
  \bibinfo{year}{2021}.
\newblock \bibinfo{title}{{FSDS}: {A} practical and fully secure document
  similarity search over encrypted data with lightweight client}.
\newblock \bibinfo{journal}{Journal of Information Security and Applications}
  \bibinfo{volume}{59}, \bibinfo{pages}{102830}.
%Type = Article
\bibitem[{Wang et~al.(2022)Wang, Sun, Wang, Liu and Chen}]{wang_achieving_2022}
\bibinfo{author}{Wang, Y.}, \bibinfo{author}{Sun, S.F.}, \bibinfo{author}{Wang,
  J.}, \bibinfo{author}{Liu, J.K.}, \bibinfo{author}{Chen, X.},
  \bibinfo{year}{2022}.
\newblock \bibinfo{title}{Achieving searchable encryption scheme with search
  pattern hidden}.
\newblock \bibinfo{journal}{IEEE Transactions on Services Computing}
  \bibinfo{volume}{15}, \bibinfo{pages}{1012--1025}.
%Type = Article
\bibitem[{Wang et~al.(2016)Wang, Wang and Chen}]{wang_secure_2016}
\bibinfo{author}{Wang, Y.}, \bibinfo{author}{Wang, J.}, \bibinfo{author}{Chen,
  X.}, \bibinfo{year}{2016}.
\newblock \bibinfo{title}{Secure searchable encryption: a survey}.
\newblock \bibinfo{journal}{Journal of Communications and Information Networks}
  \bibinfo{volume}{1}, \bibinfo{pages}{52--65}.
%Type = Inproceedings
\bibitem[{Wen et~al.(2020)Wen, Yu, Xie and Zhang}]{wen_leaf_2020}
\bibinfo{author}{Wen, R.}, \bibinfo{author}{Yu, Y.}, \bibinfo{author}{Xie, X.},
  \bibinfo{author}{Zhang, Y.}, \bibinfo{year}{2020}.
\newblock \bibinfo{title}{{LEAF}: {A} {Faster} {Secure} {Search} {Algorithm}
  via {Localization}, {Extraction}, and {Reconstruction}}, in:
  \bibinfo{booktitle}{Proceedings of the 2020 {ACM} {SIGSAC} {Conference} on
  {Computer} and {Communications} {Security}}, \bibinfo{publisher}{Association
  for Computing Machinery}, \bibinfo{address}{New York, NY, USA}. pp.
  \bibinfo{pages}{1219--1232}.
%Type = Article
\bibitem[{Whissell and Clarke(2011)}]{whissellImprovingDocumentClustering2011}
\bibinfo{author}{Whissell, J.S.}, \bibinfo{author}{Clarke, C.L.A.},
  \bibinfo{year}{2011}.
\newblock \bibinfo{title}{Improving document clustering using {Okapi} {BM25}
  feature weighting}.
\newblock \bibinfo{journal}{Information Retrieval} \bibinfo{volume}{14},
  \bibinfo{pages}{466--487}.
%Type = Article
\bibitem[{Wu et~al.(2018)Wu, Gan and Wang}]{wu_verifiable_2018}
\bibinfo{author}{Wu, D.}, \bibinfo{author}{Gan, Q.}, \bibinfo{author}{Wang,
  X.}, \bibinfo{year}{2018}.
\newblock \bibinfo{title}{Verifiable {Public} {Key} {Encryption} with {Keyword}
  {Search} {Based} on {Homomorphic} {Encryption} in {Multi}-{User} {Setting}}.
\newblock \bibinfo{journal}{IEEE Access} \bibinfo{volume}{6},
  \bibinfo{pages}{42445--42453}.
%Type = Article
\bibitem[{Xu et~al.(2021)Xu, Wang, Lu, Qu, Chen, Hu and
  Maglaras}]{10.1155/2021/5553256}
\bibinfo{author}{Xu, W.}, \bibinfo{author}{Wang, B.}, \bibinfo{author}{Lu, R.},
  \bibinfo{author}{Qu, Q.}, \bibinfo{author}{Chen, Y.}, \bibinfo{author}{Hu,
  Y.}, \bibinfo{author}{Maglaras, L.}, \bibinfo{year}{2021}.
\newblock \bibinfo{title}{Efficient private information retrieval protocol with
  homomorphically computing univariate polynomials}.
\newblock \bibinfo{journal}{Sec. and Commun. Netw.} \bibinfo{volume}{2021}.
%Type = Article
\bibitem[{Yang et~al.(2020a)Yang, Xiong and Ren}]{Yang2020DataSA}
\bibinfo{author}{Yang, P.}, \bibinfo{author}{Xiong, N.N.},
  \bibinfo{author}{Ren, J.}, \bibinfo{year}{2020}a.
\newblock \bibinfo{title}{Data security and privacy protection for cloud
  storage: A survey}.
\newblock \bibinfo{journal}{IEEE Access} \bibinfo{volume}{8},
  \bibinfo{pages}{131723--131740}.
%Type = Article
\bibitem[{Yang et~al.(2020b)Yang, Liu and Deng}]{yang_multi-user_2020}
\bibinfo{author}{Yang, Y.}, \bibinfo{author}{Liu, X.}, \bibinfo{author}{Deng,
  R.H.}, \bibinfo{year}{2020}b.
\newblock \bibinfo{title}{Multi-{User} {Multi}-{Keyword} {Rank} {Search} {Over}
  {Encrypted} {Data} in {Arbitrary} {Language}}.
\newblock \bibinfo{journal}{IEEE Transactions on Dependable and Secure
  Computing} \bibinfo{volume}{17}, \bibinfo{pages}{320--334}.
%Type = Article
\bibitem[{Yang et~al.(2020c)Yang, Liu, Deng and Weng}]{yang_flexible_2020}
\bibinfo{author}{Yang, Y.}, \bibinfo{author}{Liu, X.}, \bibinfo{author}{Deng,
  R.H.}, \bibinfo{author}{Weng, J.}, \bibinfo{year}{2020}c.
\newblock \bibinfo{title}{Flexible wildcard searchable encryption system}.
\newblock \bibinfo{journal}{IEEE Transactions on Services Computing}
  \bibinfo{volume}{13}, \bibinfo{pages}{464--477}.
%Type = Article
\bibitem[{Yin et~al.(2021)Yin, Lu, Zheng, Tang and Jiang}]{yin_achieve_2021}
\bibinfo{author}{Yin, F.}, \bibinfo{author}{Lu, R.}, \bibinfo{author}{Zheng,
  Y.}, \bibinfo{author}{Tang, X.}, \bibinfo{author}{Jiang, Q.},
  \bibinfo{year}{2021}.
\newblock \bibinfo{title}{Achieve efficient and privacy-preserving compound
  substring query over cloud}.
\newblock \bibinfo{journal}{Sec. and Commun. Netw.} \bibinfo{volume}{2021}.
%Type = Inproceedings
\bibitem[{Zhang et~al.(2020)Zhang, Shen and Huang}]{zhang_secure_2020}
\bibinfo{author}{Zhang, J.}, \bibinfo{author}{Shen, S.},
  \bibinfo{author}{Huang, D.}, \bibinfo{year}{2020}.
\newblock \bibinfo{title}{A secure ranked search model over encrypted data in
  hybrid cloud computing}, in: \bibinfo{editor}{Lu, W.}, \bibinfo{editor}{Wen,
  Q.}, \bibinfo{editor}{Zhang, Y.}, \bibinfo{editor}{Lang, B.},
  \bibinfo{editor}{Wen, W.}, \bibinfo{editor}{Yan, H.}, \bibinfo{editor}{Li,
  C.}, \bibinfo{editor}{Ding, L.}, \bibinfo{editor}{Li, R.},
  \bibinfo{editor}{Zhou, Y.} (Eds.), \bibinfo{booktitle}{Cyber Security},
  \bibinfo{publisher}{Springer Singapore}, \bibinfo{address}{Singapore}. pp.
  \bibinfo{pages}{29--36}.
%Type = Article
\bibitem[{Zhang et~al.(2018)Zhang, Xue and Liu}]{zhang_searchable_2018}
\bibinfo{author}{Zhang, R.}, \bibinfo{author}{Xue, R.}, \bibinfo{author}{Liu,
  L.}, \bibinfo{year}{2018}.
\newblock \bibinfo{title}{Searchable {Encryption} for {Healthcare} {Clouds}:
  {A} {Survey}}.
\newblock \bibinfo{journal}{IEEE Transactions on Services Computing}
  \bibinfo{volume}{11}, \bibinfo{pages}{978--996}.
%Type = Article
\bibitem[{Zheng et~al.(2021)Zheng, Zhang, Zhang and Li}]{zheng_symmetric_2021}
\bibinfo{author}{Zheng, J.}, \bibinfo{author}{Zhang, J.},
  \bibinfo{author}{Zhang, X.}, \bibinfo{author}{Li, H.}, \bibinfo{year}{2021}.
\newblock \bibinfo{title}{Symmetric searchable encryption scheme that supports
  phrase search}.
\newblock \bibinfo{journal}{Microsystem Technologies} \bibinfo{volume}{27},
  \bibinfo{pages}{1721--1727}.

\end{thebibliography}

\end{document}